\documentclass[conference]{IEEEtran}
\IEEEoverridecommandlockouts

\usepackage{mathptmx} % This is Times font
\usepackage{cite}
\usepackage{amsmath,amssymb,amsfonts}
\usepackage{algorithmic}
\usepackage{graphicx}
\usepackage{textcomp}
\usepackage{xcolor}
\usepackage{fancyhdr}
\usepackage[hyphens]{url}

\def\BibTeX{{\rm B\kern-.05em{\sc i\kern-.025em b}\kern-.08em
    T\kern-.1667em\lower.7ex\hbox{E}\kern-.125emX}}

\usepackage[keeplastbox]{flushend}
\usepackage[bookmarks=true,breaklinks=true,letterpaper=true,colorlinks,linkcolor=black,citecolor=blue,urlcolor=black]{hyperref}

\usepackage{graphicx}
\usepackage{epstopdf}
\usepackage{setspace} 
\usepackage{subfig}
\usepackage[ruled,vlined,linesnumbered]{algorithm2e}
\usepackage{url}
\usepackage{amsmath}
\usepackage{pifont}
\usepackage{fixltx2e}
\usepackage{anyfontsize}
\usepackage{kantlipsum}
\usepackage{multirow}
\usepackage[font=bf, skip=1pt]{caption}
\usepackage{array}
\usepackage{comment}
\usepackage{mwe}
\usepackage{enumitem}

\usepackage{tikz}
\usepackage{kantlipsum}

\newcommand*\whitecircled[1]{\tikz[baseline=(char.base)]{
            \node[shape=circle,fill=black,font=\bfseries,inner sep=1pt] (char) {\textcolor{white}{#1}};}}
\newcommand*\redcircled[1]{\tikz[baseline=(char.base)]{
            \node[shape=circle,fill=red,font=\bfseries,inner sep=1pt] (char) {\textcolor{white}{#1}};}}

\newcommand*\bluecircled[1]{\tikz[baseline=(char.base)]{
            \node[shape=circle,fill=blue,font=\bfseries,inner sep=1pt] (char) {\textcolor{white}{#1}};}}

\iftrue
\newcommand{\nskim}[1]{{\color{red}[\textbf{\sc nskim}: \textit{#1}]}}
\newcommand{\mj}[1]{{\color{blue}[\textbf{\sc mj}: \textit{#1}]}}
\newcommand{\jie}[1]{{\color{orange}[\textbf{\sc jie}: \textit{#1}]}}
\else
\newcommand{\nskim}[1]{}
\newcommand{\mj}[1]{}
\newcommand{\jie}[1]{}
\fi

\newcommand{\mycomment}[1]{}
\newcommand{\ignore}[1]{}
\usepackage{color}
\usepackage{soul}
\usepackage{xspace}

\begin{document}
%%%%%%%%%%%---SETME-----%%%%%%%%%%%%%
\title{Ohm-GPU: Integrating New Optical Network and Heterogeneous Memory into GPU Multi-Processors} 
% Beyond Photonic-Based Memory Channel with phantom data path
%\author{Jie Zhang and Myoungsoo Jung}
%\affiliation{
%       \institution{Computer Architecture and Memory Systems Laboratory,}
%       \country{Korea Advanced Institute of Science and Technology (KAIST)} 
%       }
%\email{http://camelab.org}
\author{
{\large Jie Zhang and Myoungsoo Jung} \\
\vspace{-2pt}
       {\large \emph{Computer Architecture and Memory Systems Laboratory,}}\\
\vspace{-2pt}
       {\normalsize{Korea Advanced Institute of Science and Technology (KAIST)}}\\
\vspace{-3pt}
	   {\large {http://camelab.org}}
       }
%%%%%%%%%%%%%%%%%%%%%%%%%%%%%%%%%%%%

%\thispagestyle{firstpage}
%\pagestyle{plain}
\maketitle

%\renewcommand{\shortauthors}{Jie Zhang and Myoungsoo Jung}

\begin{abstract}
Traditional graphics processing units (GPUs) suffer from the low memory capacity and demand for high memory bandwidth. To address these challenges, we propose \textit{Ohm-GPU}, a new optical network based heterogeneous memory design for GPUs. Specifically, Ohm-GPU can expand the memory capacity by combing a set of high-density 3D XPoint and DRAM modules as heterogeneous memory. To prevent memory channels from throttling throughput of GPU memory system, Ohm-GPU replaces the electrical lanes in the traditional memory channel with a high-performance optical network. However, the hybrid memory can introduce frequent data migrations between DRAM and 3D XPoint, which can unfortunately occupy the memory channel and increase the optical network traffic. To prevent the intensive data migrations from blocking normal memory services, Ohm-GPU revises the existing memory controller and designs a new optical network infrastructure, which enables the memory channel to serve the data migrations and memory requests, in parallel. Our evaluation results reveal that Ohm-GPU can improve the performance by 181\% and 27\%, compared to a DRAM-based GPU memory system and the baseline optical network based heterogeneous memory system, respectively.
\end{abstract}

%\begin{CCSXML}
%<ccs2012>
%<concept>
%<concept_id>10010520.10010521.10010528.10010534</concept_id>
%<concept_desc>Computer systems organization~Single instruction, multiple data</concept_desc>
%<concept_significance>300</concept_significance>
%</concept>
%<concept>
%<concept_id>10003033.10003106.10003107</concept_id>
%<concept_desc>Networks~Network on chip</concept_desc>
%<concept_significance>500</concept_significance>
%</concept>
%<concept>
%<concept_id>10010583.10010600.10010602.10010605</concept_id>
%<concept_desc>Hardware~Photonic and optical interconnect</concept_desc>
%<concept_significance>500</concept_significance>
%</concept>
%</ccs2012>
%\end{CCSXML}
%
%\ccsdesc[300]{Computer systems organization~Single instruction, multiple data}
%\ccsdesc[500]{Networks~Network on chip}
%\ccsdesc[500]{Hardware~Photonic and optical interconnect}

%\keywords{Parallel processing, graphics processing unit (GPU), optical network, Optane DC PMM, GDDR DRAM}

\section{Introduction}
\label{sec:intro}
Over the past decade, there emerged a huge number of large-scale data-intensive applications such as artificial intelligence, bigdata and cloud computing \cite{memorywall1, memorywall2, gai2016cost}. Graphics processing units (GPUs) have been widely adopted as an efficient accelerator hardware platform to speed up the execution of such applications. 
In practice, the process of large-scale applications is decomposed into a form of multiple GPU kernels. Each GPU kernel contains hundred of thousands of threads, which can be simultaneously executed by many GPU cores \cite{zhang2020zng}. 
While massively parallel computing power of a GPU can enhance data processing bandwidth, its memory system is difficult to satisfy increasing I/O demands of the large-scale applications. Specifically, DRAM faces many practical challenges to scale their technology down, and it cannot be denser due to memory retention time violations, insufficient sensing margins and low reliability issues \cite{kim2015architectural, kim2014flipping, kang2014co, nair2013archshield, yoon2013flash}. Although one can vertically stack multiple memory dies to expand the memory capacity \cite{hbm2}, stacking more memory dies will reach a near-future point where its power consumption is infeasible to be applied in GPU-like peripheral devices.
%the accumulated power consumption of massive memory dies is not affordable in GPU-like peripheral devices. 
In the meantime, bandwidth trends of the existing memory run behind those of GPU computing power. For example, while the computation throughput of the GPU cores can reach 261 TB/s (131 FP16 TFLOPS) \cite{turingGPU}, the total bandwidth of the GPU memory system is restricted by 0.7 TB/s due to the limited width of the electrical memory channels \cite{turingGPU}. %the number of electrical lanes is physically bounded by a limited number of I/O pins. Since the maximum number of electrical lanes is fixed (32-bit bus width) for DRAMs, it introduces low lane-level parallelism, which in turn constrains the maximum bandwidth of the electrical channels connecting between the memory controller and the memory media.

\begin{figure}
	\centering
	\includegraphics[width=1\linewidth]{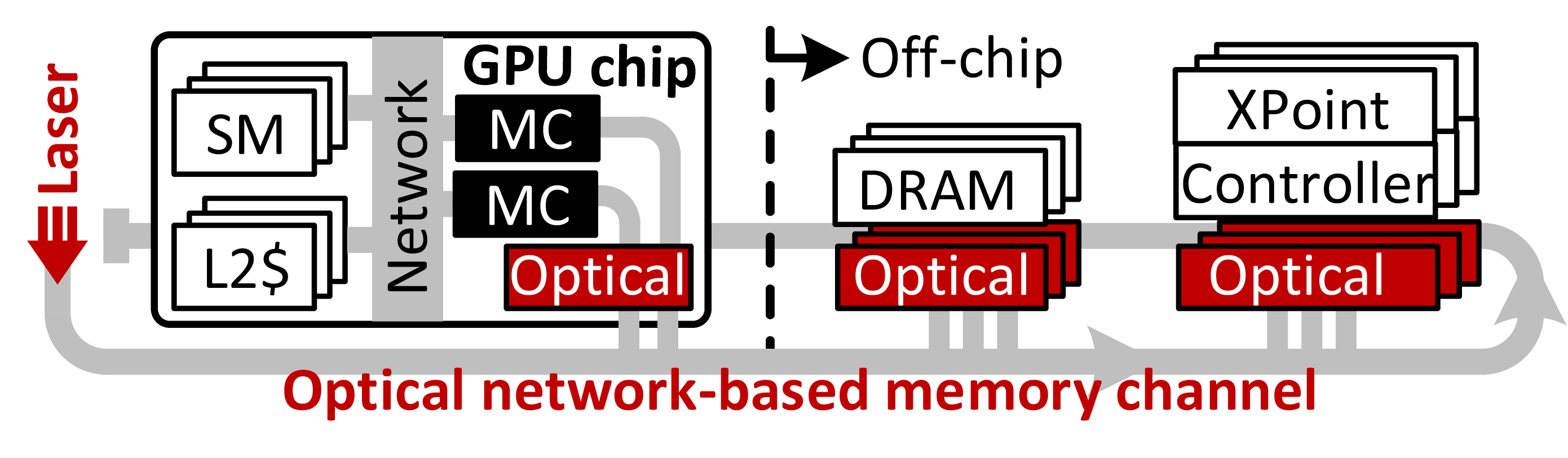}
	\vspace{-5pt}
	\caption{\label{fig:hetero}Architectural overview of Ohm-GPU.\vspace{-5pt}}
	\vspace{-5pt}
\end{figure} 

One way to improve the performance of memory channels is to parallelize data transfers with more electrical lanes. However, the number of electrical lanes is physically bounded by a limited number of I/O pins exposed by a DRAM chip. HBM2 addresses this constraint by replacing the I/O pins with through-silicon vias and microbumps \cite{siliconvia}. Even though HBM2 expands the width of a memory channel to 1024-bit, its data transfer rate per lane is much lower than that of traditional memory channel, which limits the accumulated bandwidth of HBM2 \cite{hbm2vsgddr6}.
%its manufacturing complexity prevents it from being widely employed in commercial GPUs. 
Another approach to improve the performance of memory channels is to overclock the frequency \cite{overclock}, thereby increasing the data transfer rate per electrical lane. However, this overclocking method needs to apply a higher voltage in electrical lanes to shorten the charging/discharging time of their parasitic capacitors, which can significantly increase the memory power consumption.
%The most widely adopted approach to improve the performance of memory channel is to overclock the frequency, thereby increasing the data transfer rate per electrical lane. For example, GDDR5X, an improved version of GDDR5, increases the differential write clock (WCK) from 4 GHz to 6 GHz. 
%data transfer rate per lane from 8 Gb/s to 12 Gb/s, while GDDR6, the successor of GDDR5X, further increases the data transfer rate to 16 Gb/s \cite{GDDR6}. 
%However, as the power consumption is proportional to the data transfer rate, GDDR5X increases the power consumption by 33\%, compared to GDDR5. As the data transfer rate keeps scaling to satisfy the demands of high memory bandwidth, the electrical interconnect will reach a point where its power consumption is infeasible in GPU.

On the other hand, to meet the requirements of large memory capacity and low power, one possible solution is to integrate new non-volatile memory, in particular 3D XPoint \cite{optaneDC}, into the GPU memory system as a substitute of DRAM.
%a series of prior studies proposed heterogeneous memory systems, where non-volatile memory (NVM) and small-size DRAM are employed as memory and cache, respectively \cite{wang2013exploring,mogul2009operating,park2011power,ramos2011page,dhiman2009pdram}. 
%A system can employ such heterogeneous memory system as working memory by configuring its PMEM and DRAM in either \textit{planar memory mode} or \textit{two-level memory mode}. Specifically, the planar and two-level memory modes utilize DRAM as exclusive and inclusive memory caches, respectively. 
3D XPoint, called \emph{XPoint} in this paper, has no destructive reads, and therefore, it does not require refresh power and consumes less leakage power than DRAM. In addition, the storage cores of XPoint are amenable to process scaling and formed a three-dimensional memory matrix, which in turn offers 8$\times$ greater memory spaces than the traditional DRAMs \cite{optane-size}. Nevertheless, read and write bandwidths of XPoint are respectively 4$\times$ and 6$\times$ slower than those of DRAM \cite{optane-perf}. 
%To address this low performance issue, we can employ XPoint and DRAM as memory and cache, respectively. Hot data should be migrated from XPoint-based memory to DRAM-based cache, such that memory requests can benefit from DRAM's short access latency. However, the data migration between the DRAM and XPoint frequently occupies the memory channel. The channel-level resource conflicts can stall many incoming nominal memory requests, thereby degrading GPU kernels' performance.  

In this work, we propose \textit{Ohm-GPU}, an \underline{\textbf{O}}ptical network based \underline{\textbf{h}}eterogeneous \underline{\textbf{m}}emory system for GPUs. We employ photonic waveguides \cite{shacham2008photonic,beausoleil2008nanoelectronic} as the GPU internal memory channels to replace its conventional electrical memory bus. A photonic waveguide can provide two orders of magnitude higher bandwidth and reduces the power consumption by 10$\times$ compared to a single electrical lane \cite{shacham2007design}.
To increase the memory capacity and address the latency issue of XPoint, Ohm-GPU also integrates XPoint and DRAM chips as heterogeneous memory, which can reap the benefits of DRAM's high performance while exploiting the large storage capacity offered by XPoint.
%To resolve the low performance issue of PMEM, Ohm-GPU integrates PMEM and DRAM modules as heterogeneous memory. 
%To hide the memory timing differences between DRAM and PMEM, we leverage an on-chip controller inside PMEM, called \textit{PMEM controller} that internally manages the memory transactions of 3D XPoint while exposing the same conventional memory interface that DRAM exploits. 
Figure \ref{fig:hetero} shows an architectural overview of our Ohm-GPU. The GPU-side memory controller can transfer data to off-chip DRAM and XPoint modules through an optical network-based memory channel, called \textit{optical channel}. 
%An on-chip controller inside XPoint, called \textit{XPoint controller} manages the memory transactions of XPoint internally while exposing the same conventional memory interface that DRAM exploits. 
%To reap the benefit of DRAM's high performance and exploit the large storage capacity of PMEM, this optic-heterogeneous memory subsystem needs to swap hot and cold data between PMEM and DRAM frequently \cite{dhiman2009pdram,ramos2011page}.
%In this work, we propose \textit{Ohm-GPU}, a new optical network design for the heterogeneous memory based GPU. Ohm-GPU replaces the existing electrical memory channel with optical network, which can achieve high bandwidth with low power consumption. Ohm-GPU also addresses performance degradation, imposed by the data migration on the optical channel.  
However, this baseline heterogeneous memory potentially degrades the performance due to data migrations between DRAM and XPoint.
To overcome such shortcoming, Ohm-GPU leverages unique characteristics of a silicon nano-photonic technique to create dual routes in the optical channel. The proposed dual routes can simultaneously serve the memory requests and the data migration tasks. Ohm-GPU can eliminate memory traffic generated by the data migrations on our optical channel, which can mitigate the performance degradation.
Our evaluation results show that Ohm-GPU can improve the performance by 181\% and 27\%, compared to an original DRAM-based GPU memory system and a baseline optical network based heterogeneous memory system, respectively. 
The main \textbf{contributions} of this paper can be summarized as follows:

\noindent $\bullet$ \emph{Design of optical network and heterogeneous memory in GPU.}
To improve the GPU memory throughput, Ohm-GPU integrates an optical network into the GPU memory system to replace the traditional electrical channels. As a single optical channel can connect multiple GPU memory controllers and memory chips, the memory requests in our system can compete to occupy the same optical channel, which introduces multiple channel-level I/O conflicts. To prevent the memory controllers from generating such conflicts, Ohm-GPU splits the optical channel into multiple virtual channels and assigns an individual virtual channel to each memory controller. On the other hand, to resolve the conflicts generated by memory chips, we introduce an arbitration mechanism in the optical channel, such that each memory controller can explicitly communicate with a designated memory chip at a time. To further increase GPU memory capacity, Ohm-GPU combines DRAM and XPoint as a heterogeneous memory. We introduce a novel hardware design to make conventional memory communication protocols compatible with the new optical interface. This design allows the heterogeneous memory to be easily attached to our optical network. We also integrate a logic layer into XPoint, which implements address translation, wear-levelling and transaction management. To the best of our knowledge, Ohm-GPU is the first work integrating both optical network and heterogeneous memory into a GPU.

\noindent $\bullet$ \emph{In-depth analysis of memory operational modes.}
To achieve high performance in the hybrid memory, we propose two memory operational modes in Ohm-GPU: \textit{planar memory mode} and \textit{two-level memory mode}. The planar and two-level memory modes utilize DRAM as exclusive and inclusive memory caches, respectively. While both the operational modes can hide the long latency of XPoint by migrating hot data from XPoint to DRAM, we observe that data migration between those two memories with the planar memory and two-level memory modes increase the average memory access latency by 54\% and 47\%, respectively (cf. Section \ref{sec:motivation}).
%To be precise, we perform a detailed analysis of data migration by simulating a GPU system that employs such heterogeneous memory system as global memory. The XPoint and DRAM are configured in either \textit{planar memory mode} or  \textit{two-level memory mode}. Specifically, the planar and two-level memory modes utilize DRAM as exclusive and inclusive memory caches, respectively. The data migration, imposed by the planar memory and two-level memory modes increases the average memory access latency by 54\% and 47\%, respectively, when we execute diverse memory-intensive GPU workloads (cf. Section \ref{sec:motivation} for more details).
There are two main reasons for such high data migration overhead. 
First, the data migrations are expensive as the memory controller should copy all the data to its internal buffer and redirect the data to the target memory module. Second, as our optical channel is shared by both data migration and memory requests, data migration consumes the channel resources, which should be used to serve the memory requests. %the single memory channel can be the bottleneck to serve both memory requests and data migration requests. 
We further optimize the memory system of Ohm-GPU to mitigate the performance overhead of data migration.

\begin{figure}
	\centering
	\includegraphics[width=1\linewidth]{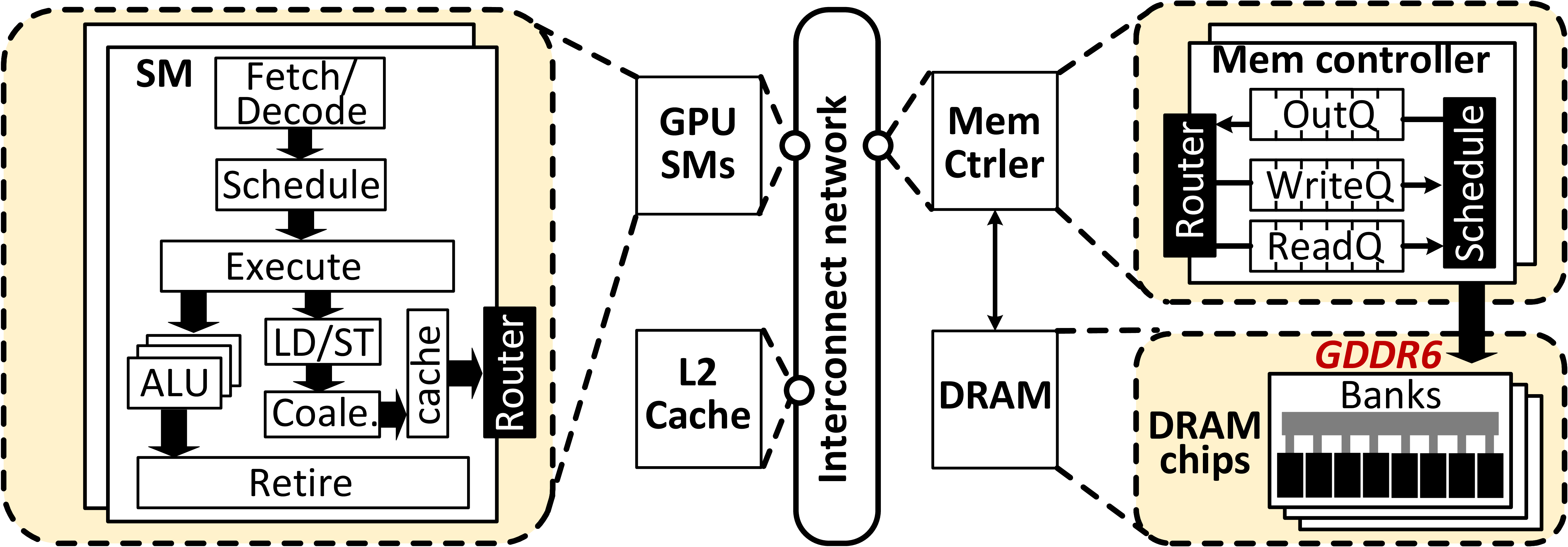}
	\vspace{-5pt}
	\caption{\label{fig:baseline_gpu}Overview of baseline GPU architecture.}
	\vspace{-15pt}
\end{figure}

\noindent $\bullet$ \emph{Memory system optimization for low memory traffic.}
To reduce bandwidth costs of the traditional data copy operations, Ohm-GPU enables XPoint controllers to directly migrate data between DRAM and XPoint. To prevent the memory controller and the XPoint controller from competing to access the same DRAM, Ohm-GPU implements a conflict detection mechanism in the memory controllers, which can detect the potential conflicts before scheduling the memory requests and data migration requests. To make the memory requests fully exploit the optical channel, we create dual routes in the same optical channel to simultaneously serve the memory requests and the data migration tasks. Our new design requires minor optical hardware costs and does not increase the total energy consumption of the target memory system.

\section{Background}
\label{sec:background}
\begin{figure}
\centering
\vspace{-10pt}
\subfloat[GPU-SSD integrated system.]{\label{fig:exe_brkdown}\rotatebox{0}{\includegraphics[width=0.48\linewidth]{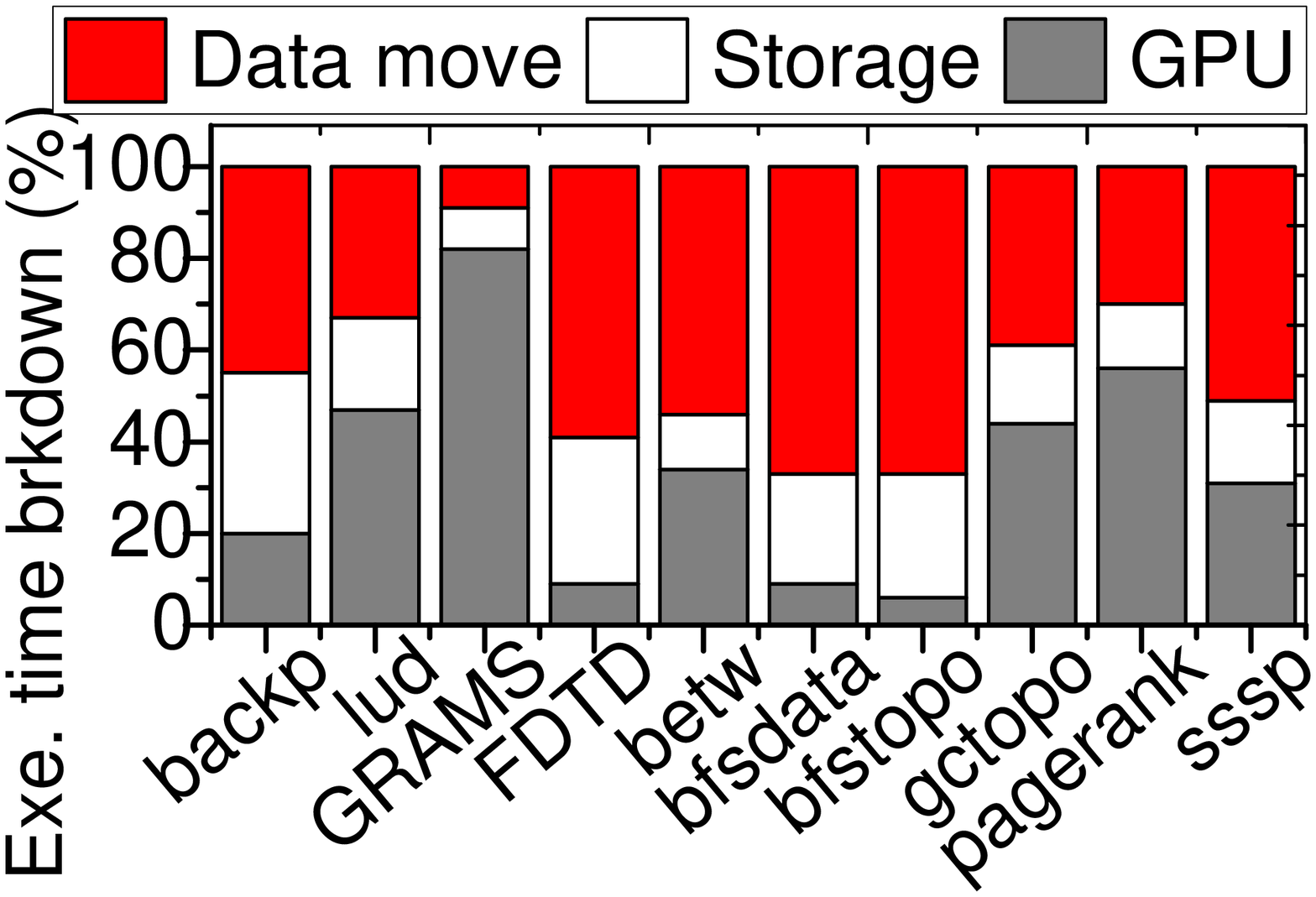}}}
%\hspace{1pt}
\subfloat[GPU memory subsystem.]{\label{fig:dram_brkdown}\rotatebox{0}{\includegraphics[width=0.52\linewidth]{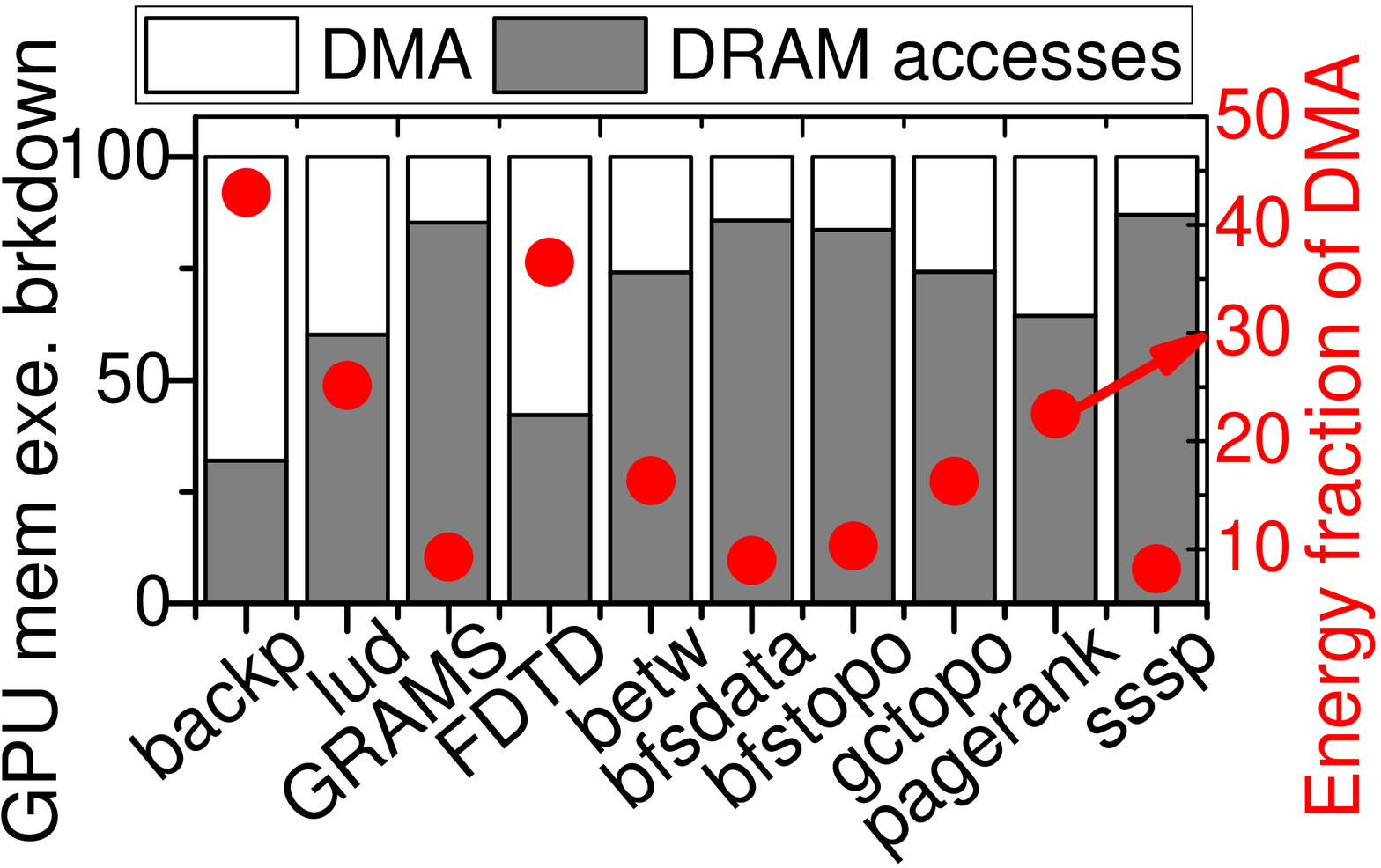}}}
%\vspace{-10pt}
\caption{\label{fig:brkdown}Breakdown analysis of executing GPU apps.\vspace{-5pt}}
%\vspace{-5pt}
\end{figure}

%In this section, we describe the details of the optic-heterogeneous memory, including the optical network design and the heterogeneous memory control logic design.
\subsection{Baseline GPU Architecture}
Figure \ref{fig:baseline_gpu} shows a baseline GPU architecture, which is similar to the real-world GPU products \cite{nvidia2009nvidia, nvidia2012nvidia, turingGPU, gv100}. Specifically, the baseline GPU consists of multiple streaming multiprocessors (SMs), shared L2 cache and memory controllers, all of which are connected through an interconnect network. Within the SMs, a group of 32 threads, called \emph{warp} \cite{nvidia2009nvidia}, are executed in a lockstep. During the execution, a set of instructions for each warp is fetched from the underlying GPU memory. The instructions are then decoded and stored in the register files. Afterwards, the warp scheduler schedules the warps to execute. Arithmetic instructions are executed by ALUs, while load/store instructions generate memory requests. 
SMs firstly try to find out data associated with the memory requests from L1D cache. If L1D cache misses, the requests will be forwarded to the shared L2 cache via the interconnect network. If L2 cache also misses, the requests will be sent to the memory controller.
%The coalescing units will firstly coalesce the 32-bit memory requests into fewer but larger memory requests and forwards the requests to the L1D cache. 
A traditional GPU memory controller in practice buffers and schedules incoming memory requests \cite{zhang2020zng}. For each request, the memory controller issues the memory transactions with DRAM via GDDR6 protocol. 

\begin{figure}
	\centering
	\includegraphics[width=1\linewidth]{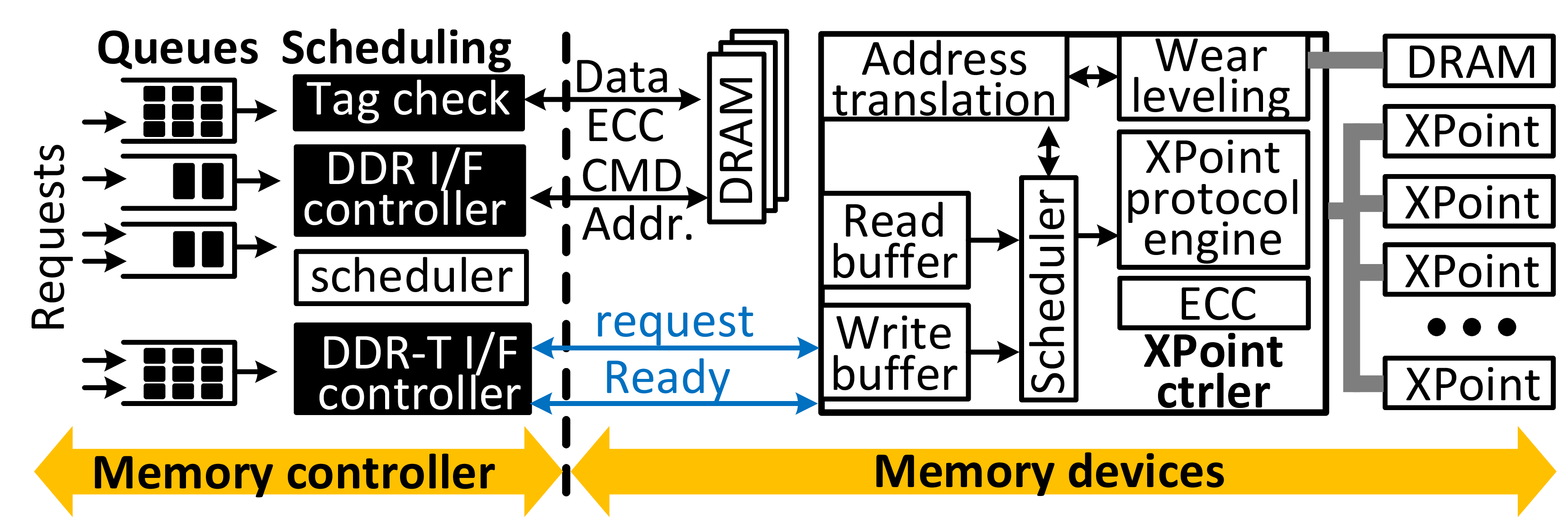}
	%\vspace{-15pt}
	\caption{\label{fig:mc}Memory controller in heterogeneous memory.\vspace{-5pt}}
	\vspace{-5pt}
\end{figure}

\subsection{Challenges in GPU Memory System}
The existing GPU memory system becomes the performance bottleneck of executing large-scale applications in the GPU due to its low capacity, limited throughput, and high energy consumption \cite{zhang2015nvmmu,zhang2020zng,zhang2019flashgpu}. As the GPU on-chip DRAM cannot accommodate all data sets of large-scale applications, the data often require being loaded/stored from/to an external storage. To be precise, we evaluated a computing system that integrates a high-performance GPU \cite{nvidia2012nvidia} and an SSD \cite{ZSSD} together. Figure \ref{fig:exe_brkdown} shows a breakdown analysis to execute different GPU applications \cite{che2009rodinia,pouchet2012polybench,nai2015graphbig} on our testbed system. The storage access delay and data transfers between the GPU and SSD account for 21\% and 45\% of the total execution time, on average, respectively. This data movement overhead takes a time longer than the GPU computing time itself by 2.3$\times$, on average. We also analyze the impact of DMA and DRAM accesses on the GPU memory system in terms of execution time and energy consumption, and the results are shown in Figure \ref{fig:dram_brkdown}. Transferring data via electrical memory channels (i.e., DMA) degrades the performance of the GPU memory system by 31\% and 19\% in terms of execution time and energy consumption, respectively. 

%\begin{figure}
%\centering
%\subfloat[Optical network-based memory channel.]{\label{fig:optical_background}\rotatebox{0}{\includegraphics[width=0.74\linewidth]{figs/optical_background1}}}
%\hspace{1pt}
%\subfloat[Modulator.]{\label{fig:mrr}\rotatebox{0}{\includegraphics[width=0.24\linewidth]{figs/mrr1}}}
%\caption{\label{fig:bakground}Overview of optical network-based memory channel and photonic modulator.}
%\vspace{-10pt}
%\end{figure}

%\begin{figure}
%	\centering
%	\includegraphics[width=1\linewidth]{figs/optical_background.eps}
%	\vspace{-20pt}
%	\caption{\label{fig:optical_background}The optical channel.\vspace{-15pt}}
%	%\vspace{-5pt}
%\end{figure}
%
%\begin{figure}
%	\centering
%	\includegraphics[width=1\linewidth]{figs/mrr.eps}
%	\vspace{-15pt}
%	\caption{\label{fig:mrr}An example of how photonic modulator works and different types of active MRRs.\vspace{-15pt}}
%	\vspace{-5pt}
%\end{figure}

%The CMOS-compatible integrated silicon-photonic technology makes it possible to build a optical network to replace the existing electrical memory channel \cite{shacham2007design, beausoleil2008nanoelectronic}. Compared to the electrical memory channel, the dense wavelength-division multiplexing (DWDM) technique in the optical network allows different wavelengths in a single waveguide to carry multiple data streams, which can deliver two orders of magnitude higher bandwidth.

\subsection{Heterogeneous Memory System}
\label{sec:mcarch}
Combining DRAM and 3D XPoint (\emph{XPoint}) as a heterogeneous memory system is considered as a good solution to take advantage of DRAM's high-performance and XPoint's large-capacity in parallel. Figure \ref{fig:mc} shows the details of a heterogeneous memory system proposed by prior work \cite{ramanujan2014dynamic, nale2019memory}. One can observe from the figure that the heterogeneous memory system places both DRAM and XPoint together to share the same memory channel for data transfers. However, DRAM and XPoint require different memory control logic to manage their communication protocols and process memory transactions. To address this, the heterogeneous memory system includes a customized memory controller and a XPoint controller to handle the communication protocols \cite{ramanujan2014dynamic, nale2019memory}. %In particular, the memory controller schedules the memory requests and interacts with the PMEM controller via a communication protocol to access data.

% non-deterministic latency->async communication, limited lifetime->wear leveling, address translation
%PMEM, as a new type of persistent memory, exhibits different characteristics from the traditional DRAM. Unfortunately, it makes the PMEM difficult to be directly attached to the DRAM DIMM slots. Specifically, as the storage cores of PMEM, referred as 3D Xpoint, takes variable latencies for cell sensing and programming, the memory access latencies of PMEM are non-deterministic, which is not compatible to the deterministic timing protocol such as DDR4. In addition, the 3D Xpoint has limited lifetime, which can easily wear out due to the intensive memory accesses. Therefore, PMEM requires additional protection for better endurance.

\noindent \textbf{XPoint controller.} Unfortunately, the memory controller cannot directly communicate with XPoint due to two root causes: 1) XPoint operates in a different clock frequency from the memory channel; 2) Xpoint has a limited lifetime, which can easily wear-out if there exist intensive memory accesses. A XPoint controller is deployed between the memory controller and XPoint to handle the different clock frequency and the wear-out issues of XPoint, as shown in Figure \ref{fig:mc}. Specifically, to address the asymmetric frequency issue, the XPoint controller employs read and persistent write buffers, which temporarily store the incoming memory requests and data. The XPoint controller then processes the memory requests asynchronously and leverages a XPoint protocol engine to access data from XPoint.  
To improve the endurance of XPoint, the XPoint controller also employs address translation and wear-levelling algorithms, which are similar to the reliability management of flash-based SSD controllers \cite{micheloni2013inside,kim2002space,gupta2009dftl,chang2007efficient}. The ``metadata'' of these algorithms (e.g., address mapping table) are stored in an external DRAM buffer \cite{optane-test}.
%To improve the endurance of 3D XPoint, the address translation unit remaps the memory requests to point to different physical addresses of 3D XPoint. 
%After address translation, the XPoint protocol engine then interacts with 3D XPoint to serve the memory commands.

\noindent \textbf{Memory controller.}  
%Similar to traditional DRAM controller, the memory controller in the heterogeneous memory uses the same request management functions, such as request buffering, memory mapping and scheduling. As shown in the figure, the incoming memory requests are firstly buffered in the request queues (\textit{request buffering}). The addresses of the buffered requests are then decoded to find the target memory. Based on the metadata information, the memory requests are converted into memory commands, which are moved to the queues corresponding to the memory media (\textit{memory mapping}).    
%Afterwards, the memory controller schedules the memory commands based on a specific request scheduling policy (e.g., FCFS and FRFCFS) and forwards the selected commands to an interface controller. However, 
In contrast to the DRAM controller, the memory controller employs two different protocols, DDR and DDR-T \cite{ramanujan2014dynamic}, to communicate with DRAM and the XPoint controller, respectively. This is because the memory access latencies of XPoint are non-deterministic and incompatible with the deterministic memory timing protocol such as DDR;  
DDR-T is an asynchronous communication protocol. Specifically, after sending commands to the XPoint controller, the memory controller switches to process other memory requests. Once the memory controller gets any response message from XPoint controller, it turns to receive the data over memory channel.
%Specifically, DDR-T prohibits the memory controller to directly fetch the data from PMEM. Instead, PMEM needs to inform the memory controller if the data is ready. Once it receives the confirmation message from the memory controller, PMEM can send target data and completion message to the memory controller.

\begin{figure}
	\centering
	\includegraphics[width=1\linewidth]{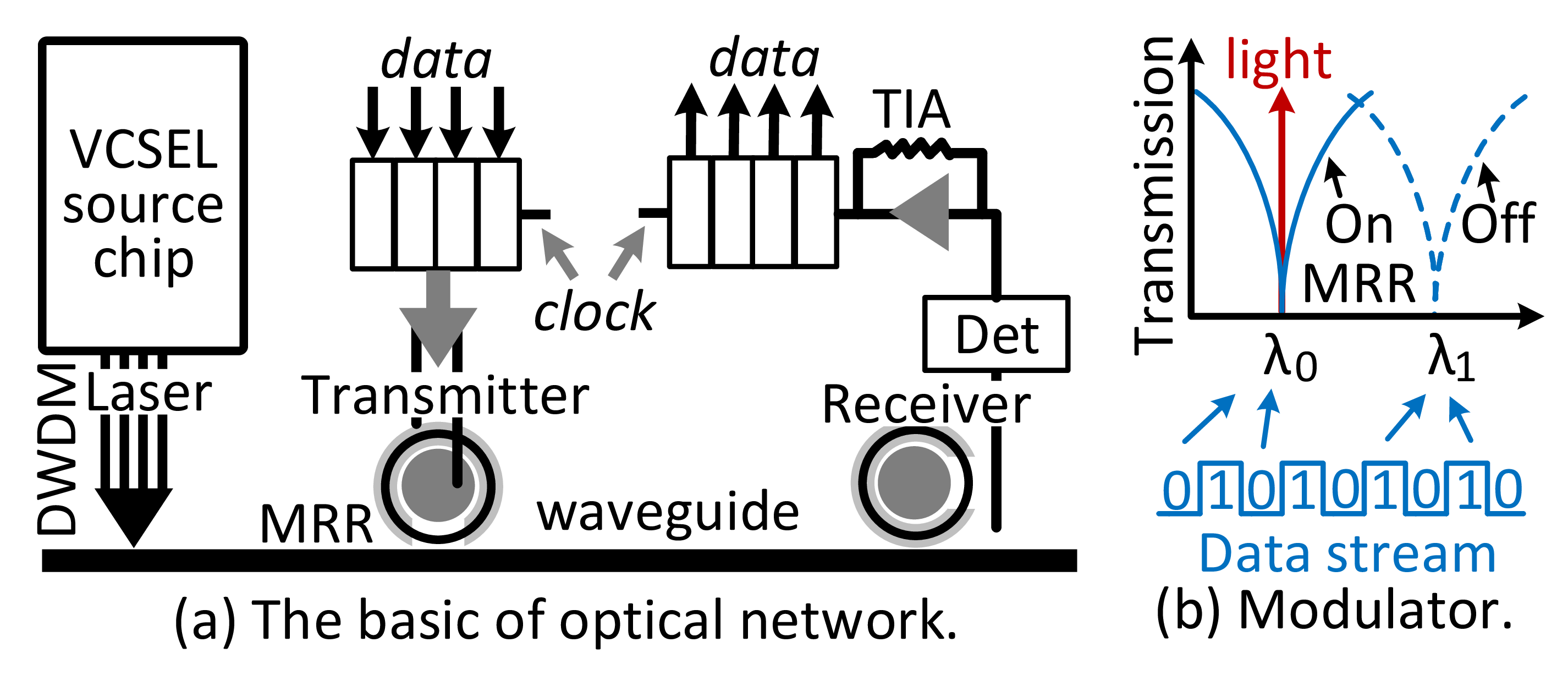}
	%\vspace{-15pt}
	\caption{\label{fig:optic_basic}Details of optical network and modulator.\vspace{-5pt}}
	\vspace{-5pt}
\end{figure}

\begin{figure*}
	\centering
	\includegraphics[width=1\linewidth]{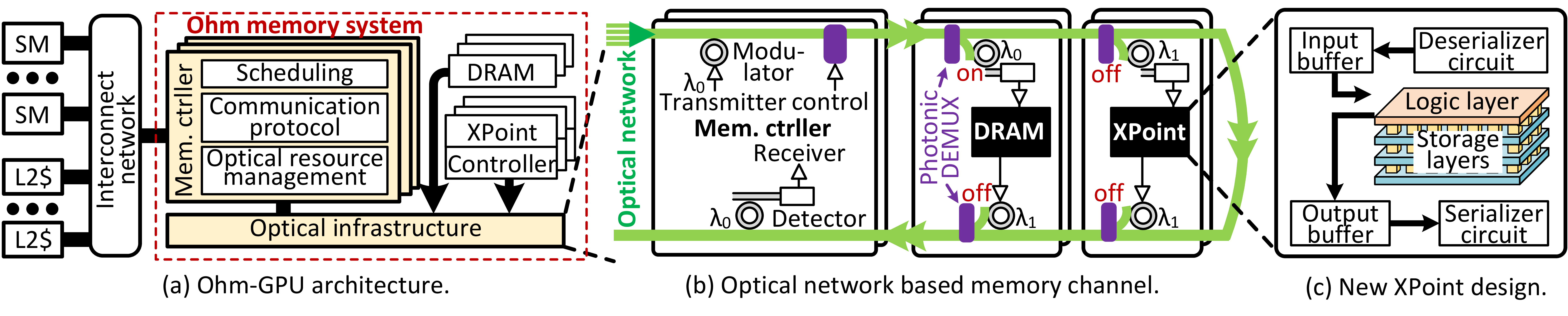}
	%\vspace{-15pt}
	\caption{\label{fig:background}Overview of Ohm-GPU architecture with an Ohm memory system.\vspace{-10pt}}
	%\vspace{-10pt}
\end{figure*}

\subsection{Optical Network}
\label{sec:optical}
\noindent \textbf{Optical channel.}  
Silicon nano-photonic technologies enable an optical network to replace the existing electrical memory channel \cite{shacham2007design, beausoleil2008nanoelectronic}. The optical channel can be an ideal candidate to connect between memory controllers and multiple memory chips as it improves the interconnect bandwidth by two orders of magnitude and reduces the power consumption of an electrical channel by 10$\times$ \cite{li2013exploring}. Figure \ref{fig:optic_basic}a shows a typical design of an optical channel \cite{bergman2014photonic}. In contrast to the electrical memory channels, the optical channel uses laser light of multiple wavelengths to carry different data streams from multiple transmitters to receivers. These laser lights traverse through a transmission medium, referred to as \textit{waveguide} \cite{bergman2014photonic}. 
%The data transfer in the optical channel also requires laser sources, photonic modulators and photonic detectors \cite{bergman2014photonic}. 
The external vertical-cavity surface-emitting laser (VCSEL) sources generate lights of different wavelengths and inject the lights into a single waveguide by using dense wavelength-division multiplexing (DWDM) \cite{bergman2014photonic}. To send a message via the optical channel, each transmitter employs a micro-ring resonator as a photonic modulator to modulate its electrical data on the light generated by the laser source \cite{bergman2014photonic}. On the other hand, the receivers can get the messages by using a micro-ring resonator as a photonic detector to couple and absorb the incoming laser lights. The receiver then converts the light bits into electrical signals, which will be forwarded to the controllers or memories.

\noindent \textbf{Micro-ring resonator.} In this work, we adopt an \textit{active micro-ring resonator} (MRR) to implement photonic modulators and detectors \cite{bahadori2016comprehensive}. MRR is a closed loop of transmission medium, which can couple and absorb the light of a wavelength by tuning the resonance wavelength.
%a MRR can be easily integrated in the memory controller and memory devices \cite{xu200712}, due to its small size (i.e., 1.5 $\sim$ 4 um diameter). 
A photonic modulator leverages MRR to modulate the laser lights.
Figure \ref{fig:optic_basic}b shows a transmission power distribution, generated by an MRR-based photonic modulator. 
When MRR is tuned to the resonance wavelength $\lambda_{0}$, the photonic modulator fully couples the light of wavelength $\lambda_{0}$, and thus, the transmission power of wavelength $\lambda_{0}$ in a waveguide is low, denoted by logic ``0''. Otherwise, when MRR is tuned to resonance wavelength $\lambda_{1}$, the light of wavelength $\lambda_{0}$ directly passes the MRR. At this time, the transmission power of wavelength $\lambda_{0}$ in the waveguide is high, denoted by logic ``1''. The photonic detector also leverages MRR to couple/absorb the light. It detects the data by sensing the light strength. Note that a laser light can be used by a single pair of photonic modulator and detector. This is because, once the light of a wavelength is fully coupled in an MRR, other MRRs cannot absorb or modulate data in the same light. Due to this, incoming nominal memory requests and data migration have to be serialized in the optical channel. %Thus, the optical channel operates as a unidirectional point-to-point interconnection \cite{li2013exploring}. The way to tune the resonant wavelength of an active MRR is to change its refractive index \cite{levy2011harmonic}. Figure \ref{fig:mrr}b and \ref{fig:mrr}c show the design details of electrical-tuned MRR and optical pulse-tuned MRR, respectively. As shown in the figure, the electrical-tuned approach changes the refractive index by applying the external electrical voltage to inject current in the resonator. In contrast, the optical pulse-tuned approach injects the free carriers to change the refractive index \cite{almeida2004all}. The electrical-tuned approach fits for photonic modulator, as it can directly turn electrical data stream into different resonance wavelengths. Meanwhile, we adopt the optical pulse-tuned approach as photonic detector, due to its short tuning delay (40 ps) \cite{ibrahim2004photonic}.

\section{High-level View of Ohm-GPU}
Figure \ref{fig:background}a shows an overview of our baseline Ohm-GPU design. Compared to the traditional GPU (cf. Figure \ref{fig:baseline_gpu}), Ohm-GPU integrates a new memory system, called \textit{Ohm memory system}, to replace the existing DRAM-based GPU memory system. Specifically, the Ohm memory system employs DRAM and XPoint as a heterogeneous memory to increase the memory capacity while maintaining high performance. DRAM in Ohm-GPU also accommodates write-intensive data, which can significantly reduce the number of writes on XPoint, thereby extending the lifetime of XPoint. To improve the bandwidth and energy consumption behaviors of the memory system, Ohm-GPU also integrates an optical infrastructure, which jointly connects the memory controllers and the memory devices. 
%The memory controllers are modified to support different communication protocols of DRAM and XPoint. In addition, they are responsible for managing the optical infrastructure.
\subsection{Ohm-GPU Architectural Design}
\noindent \textbf{Optical infrastructure for Ohm-GPU.}
%\noindent \textbf{Channel arbitration.}  
Figure \ref{fig:background}b shows a high-level overview of our optical infrastructure design. %which can be integrated into the GPU interconnect network. 
An optical channel replaces hundreds of electrical lanes to connect between the GPU memory controllers and memory devices. 
%Traditional design usually employs dual optical channels to enable duplex communications between two endpoints. Since the GDDR and DDR-T communication protocols constrains a single memory controller or device to transfer data at a time, we only employ a single optical channel to enable half-duplex communication. Therefore, the memory controllers and devices employ both photonic transmitters and receivers to connect to the single optical channel.
Directly attaching multiple memory controllers to a single optical channel can introduce channel-level conflicts, as these memory controllers can compete to occupy the same optical channel. To address this challenge, we statically split the optical channel into multiple virtual channels and assign a dedicated virtual channel to each memory controller. Our ``key insight'' is that an optical channel transfers different data streams in multiple wavelengths by using the wavelength-division multiplexing. Therefore, we can split the available wavelengths to compose different virtual channels.
While the virtual channels ensure no channel conflicts among all memory controllers, the transmitters and receivers of massive memory devices may compete to occupy a virtual channel.
%Another challenge is that, as multiple multiple memory devices exist in the memory system, their transmitters and receivers may compete to occupy the single optical channel. 
To address this, we leverage the control logic and photonic demultiplexers proposed in \cite{li2013exploring} to arbitrate the competition of an optical channel usage, as shown in Figure \ref{fig:background}b. Specifically, if a memory device is ready to serve memory requests, the photonic demultiplexer sets up the communication between the memory controller and the memory device (cf. Figure \ref{fig:background}b) by enabling the photonic detector of this memory device. Meanwhile, the photonic detectors of the remaining memory devices are disabled to yield the optical channel.

\noindent \textbf{Integration of DRAM and XPoint.}
%Figure \ref{fig:background}a shows a design overview for our baseline optic-heterogeneous memory that modifies a photonic-enabled DRAM \cite{li2013exploring}. 
Unfortunately, DRAM and XPoint cannot be directly attached to the optical channel via the photonic transmitters and receivers. There are two reasons. Firstly, the command, address, and data are simultaneously accessed in the memory devices, while all the data are serialized in the optical channel. Secondly, XPoint requires assistance of a XPoint controller to enable ECC, manage the endurance of XPoint, and control its I/O transactions. Figure \ref{fig:background}c shows our designs to address such challenges. To make the XPoint and DRAM compatible with the optical channel, we employ a SerDes circuit \cite{aoyama20033} to transform data between serial and parallel I/Os. We also employ a small size of registers (i.e., 16KB) in front of the memory devices to buffer the data from the optical channel. To integrate XPoint in the optical channel, one simple solution is to employ a XPoint controller for each XPoint. However, having multiple XPoint controllers can increase the area cost, which is critical as the limited GPU space is concerned. In addition, the XPoint controller requires a DRAM buffer to store the metadata of address translation. Since XPoint stacks its storage cores into multiple layers, we can save the area cost by integrating the XPoint controller in XPoint as a logic layer, which is adopted by several prior logic-in-memory designs\cite{pawlowski2011hybrid,shulaker2014monolithic,zhu2013accelerating}. In addition, we implement a simple wear-levelling scheme inspired by Start-Gap \cite{qureshi2009enhancing}, which periodically shifts the logical address of incoming I/Os rather than maintaining a huge mapping table in DRAM buffer \cite{zhou2009durable}. Thus, our XPoint controller design can fully eliminate the usage of the DRAM buffer.
%Although we replace the electrical lanes with a photonic waveguide, there is no obstacle to exploit the traditional GDDR6 and DDR-T protocols. Thus, the GPU memory controllers in our Ohm-GPU employs different communication protocol engines to interact with DRAM and PMEM properly. 

\subsection{Operational Modes of Ohm Memory}
\label{sec:operational}

%\subsection{Case Study of Optic-Heterogeneous Memory}
%\label{sec:operational}
In our design, the Ohm memory system can be organized in two operational modes to meet different purposes.

%\noindent \textbf{System persistency mode.}
%In a persistent memory system, a large battery is provisioned to back up the data from DRAM to PMEM, in cases of a power outage \cite{narayanan2012whole, kateja2017viyojit}. \newedit{However, the battery capacity may not be sufficient to back up the whole DRAM modules \cite{kateja2017viyojit}.
%To address this, the OS actively detects the number of dirty pages in DRAM. If the current battery budget is going to be insufficient to back up the increasing dirty pages in DRAM, the OS actively backs up the dirty pages until the remaining dirty pages do not exceed the battery budget.} However, the backup procedure can block the execution of user applications, when data migration and memory requests compete for the memory resources, as shown in Figure \ref{fig:motiv}a. Specifically, when OS generates many memory transactions to migrate the dirty pages from DRAM to PMEM, read/write requests and the corresponding data can fully occupy the memory channel. At this juncture, the DRAM cannot serve the memory requests from the user application until the memory channel is free to use.

\begin{figure}	
	\centering
	\includegraphics[width=1\linewidth]{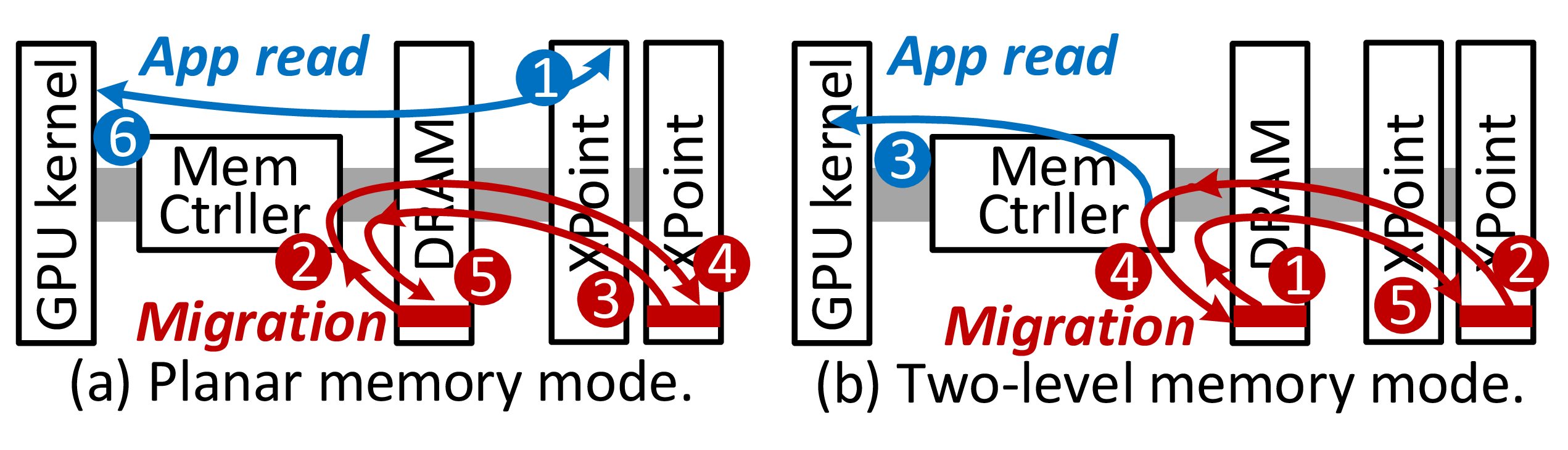}
	%\vspace{-15pt}
	\caption{\label{fig:motiv}Data movement in heterogeneous memory.\vspace{-5pt}}
	\vspace{-5pt}
\end{figure}

\noindent \textbf{Planar memory mode.}
To maximize the capacity of our heterogeneous memory, DRAM and XPoint can be configured as a planar memory, incarnating a unified address space. However, as XPoint is directly exposed to GPU kernels in this memory mode, GPU kernels suffer from the long NVM read and write delays. To hide such latency, one may use DRAM as prefetching caches by loading hot data from XPoint in advance and serving them directly from DRAM. 
Unfortunately, each data movement introduces significant OS overhead, including changing page table entries and shooting down translation lookaside buffers. To address this challenge, 
We adopt an OS-transparent data migration design inspired by \cite{wang2016duang}. Specifically, the entire memory space is split into multiple groups, each containing a DRAM page and a few XPoint pages based on the capacity ratio of DRAM and XPoint. If a XPoint page experiences intensive memory accesses, the XPoint page is considered as hot, and thus, the data of the XPoint page is swapped with the data of a DRAM page (cf. \redcircled{2}\redcircled{3}\redcircled{4}\redcircled{5} in Figure \ref{fig:motiv}a). The memory controllers record the mapping information of logical address and data location in a simplified mapping table. When serving memory requests, the memory controllers look up the table for the target data. Note that during this \textit{swap}, the memory controller can still issue memory requests coming from the GPU kernels to access DRAM/XPoint, which are not busy (\bluecircled{1}). However, the data migration generated by the swap procedure blocks the memory channel, which postpones the data response for the GPU kernels with a significantly long delay (\bluecircled{6}).

\noindent \textbf{Two-level memory mode.} 
DRAM in a two-level memory mode is configured as an inclusive cache of XPoint \cite{yang2020empirical,dhiman2009pdram,park2018bibim}. 
We adopted a typical design of memory controllers \cite{nale2019memory,yang2020empirical} to implement the two-level memory mode, and it employs a tag check module to examine if data presents in DRAM (cf. Figure \ref{fig:mc}). 
Specifically, as shown in Figure \ref{fig:motiv}b, when a memory request arrives at the memory controller, its address is decoded into an index, a tag, and an offset. The memory controller scans the DRAM cache, based on the index of the memory request (\redcircled{1}). The tag check module then compares the tag of the memory request with the metadata of the target DRAM cache-line. If matched, the memory controller directly services the memory request with the data fetched from DRAM. Otherwise, the memory controller accesses the data from XPoint based on the target address of the memory request (\redcircled{2}). The memory controller then serves the memory requests with the fetched data (\bluecircled{3}). Note that it is impractical to build a large tag array within the controller as it requires a high space to store the metadata of the entire DRAM cache. 
Instead, we configure DRAM as a direct-map cache, which can reduce a large tag size to a few bits. Thanks to the reduced tag size, the memory controller can store the small metadata (including 1 valid bit, 1 dirty bit, and 3$\sim$6 tag bits) along with ECC in the ECC region of each DRAM cache-line \cite{nale2019memory}. This new design can eliminate the space cost of the tag array. In addition, while traditional DRAM cache fetches the metadata and data via two separate memory accesses, the memory controller in our new design can fetch the data, ECC, and corresponding metadata from a single DRAM cache-line. This can eliminate potential overhead of the metadata accesses. Nevertheless, the controller is yet frequently involved in migrating data between DRAM and XPoint (\redcircled{4}\redcircled{5}), which can impose migration overhead.

\section{Designs for Migration Reduction}
\label{sec:highlevelview}
%The design goal of our Phantom is to address the negative impact of data migration in the heterogeneous memory by being aware of the characteristics of the optical network. To achieve this goal, we specify four techniques: \textit{1) introducing new PMEM functions to simplify the data movement;} \textit{2) designing a new type of MRR-based photonic modulators and detectors to support the new PMEM functions in optical channel;} \textit{3) optimizing the existing optical channel to reduce the overhead of embedding the new photonic modulators and detectors;}  
%and \textit{4) proposing specific optimizations for different operational modes.} 
%In this section, we will quantitatively analyze the impact of data migrations, provide the overview of the proposed memory system, and describe design details of each implementation.

\begin{figure}
	\centering
	\includegraphics[width=1\linewidth]{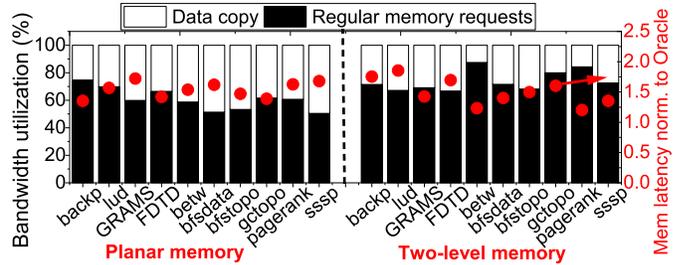}
	%\vspace{-10pt}
	\caption{\label{fig:motiv1}Memory bandwidth and latency analysis.\vspace{-5pt}}
	\vspace{-5pt}
\end{figure}

\subsection{Challenges of Ohm Memory System}
\label{sec:motivation}
We perform a simulation-based study of the two heterogeneous memory modes to measure the data migration overheads observed in the baseline Ohm memory system. We measure the effective memory bandwidth (consumed by different GPU workloads) and the wasted memory bandwidth (used for the data migration) of diverse benchmarks \cite{che2009rodinia,pouchet2012polybench,nai2015graphbig}. In addition, we compare average memory access latencies of our baseline Ohm memory system (\texttt{baseline}) with an oracle Ohm memory system (\texttt{Oracle}), which allocates a dedicated optical channel for the data migration. Figure \ref{fig:motiv1} shows each fraction of effective and wasted memory bandwidths. Our evaluation shows that the data migration in the planar memory mode accounts for 39\% of the total memory bandwidth, on average. Similarly, in the two-level memory mode, the data migration between DRAM and XPoint overall accounts for 26\% of total memory bandwidth. As a result, compared to \texttt{Oracle}, the data migration of \texttt{baseline} increases the average memory access latencies by 54\% and 47\% in the planar and two-level memory modes, on average, respectively.

%To sum up, the root causes of data migration overhead are two aspects: first, migrating a single data block between DRAM and PMEM generates more memory traffic than a single memory request; second, data migration competes with memory requests for a optical channel.

%\begin{figure}
%	\centering
%	\includegraphics[width=1\linewidth]{figs/timing_overall.eps}
%	\vspace{-15pt}
%	\caption{\label{fig:timing_overall}Ideal memory system without migration overhead.\vspace{-5pt}}
%	\vspace{-10pt}
%\end{figure}

\subsection{Memory System Design to Remove Migration Overhead}
%\noindent \textbf{Overall design.} Figure \ref{fig:timing_overall} shows the timing diagram of our target memory system design, which can hide the data migration penalties. 
%As shown in the figure, in addition to the data route between the memory controller and memory devices, Ohm-GPU can simultaneously create an independent route in the same optical channel to service data migration. This design is referred to as \emph{dual routes}.
%%The extra data path is called as \emph{phantom} path. System persistency mode can benefit from phantom path in two ways: first, dirty data blocks are simultaneously backed up via phantom path, when user applications send write requests to DRAM; second, the backup procedure generates less memory traffic by sending dirty pages from DRAM to PMEM via phantom path. 
%%The system persistency mode can benefit from our phantom path (cf. Section \ref{sec:operational}). The backup procedure generates less memory traffic by sending dirty pages from DRAM to PMEM via our phantom path.   
%In the planar memory mode, Ohm-GPU can in parallel migrate data using the independent route, while serving the memory requests of GPU kernels via the original data route. In the two-level memory mode, Ohm-GPU can reduce the memory traffic if there is a DRAM cache miss, as data can be transferred between DRAM and XPoint through the independent route. 

\noindent \textbf{Overall design.} To hide the data migration penalties, we propose a design of \emph{dual routes} that can serve the memory requests and data migration in parallel. 
Specifically, we design a new optical network infrastructure to create two routes in the same optical channel: 1) a \emph{data route} connects the memory controller and memory devices together and 2) a \emph{memory route} connects between two memory devices. While processing memory requests via the data route of an optical channel, Ohm-GPU can simultaneously perform the data migration via the memory route in the same optical channel. 
Since the dual routes are unfortunately not supported by the existing memory system, we also introduce a few new memory system designs to reap the benefits of our dual routes. 
%We will describe a new optical infrastructure design for dual routes in Section \ref{} and explain the memory system design shortly. 

\begin{figure}
	\centering
	\includegraphics[width=1\linewidth]{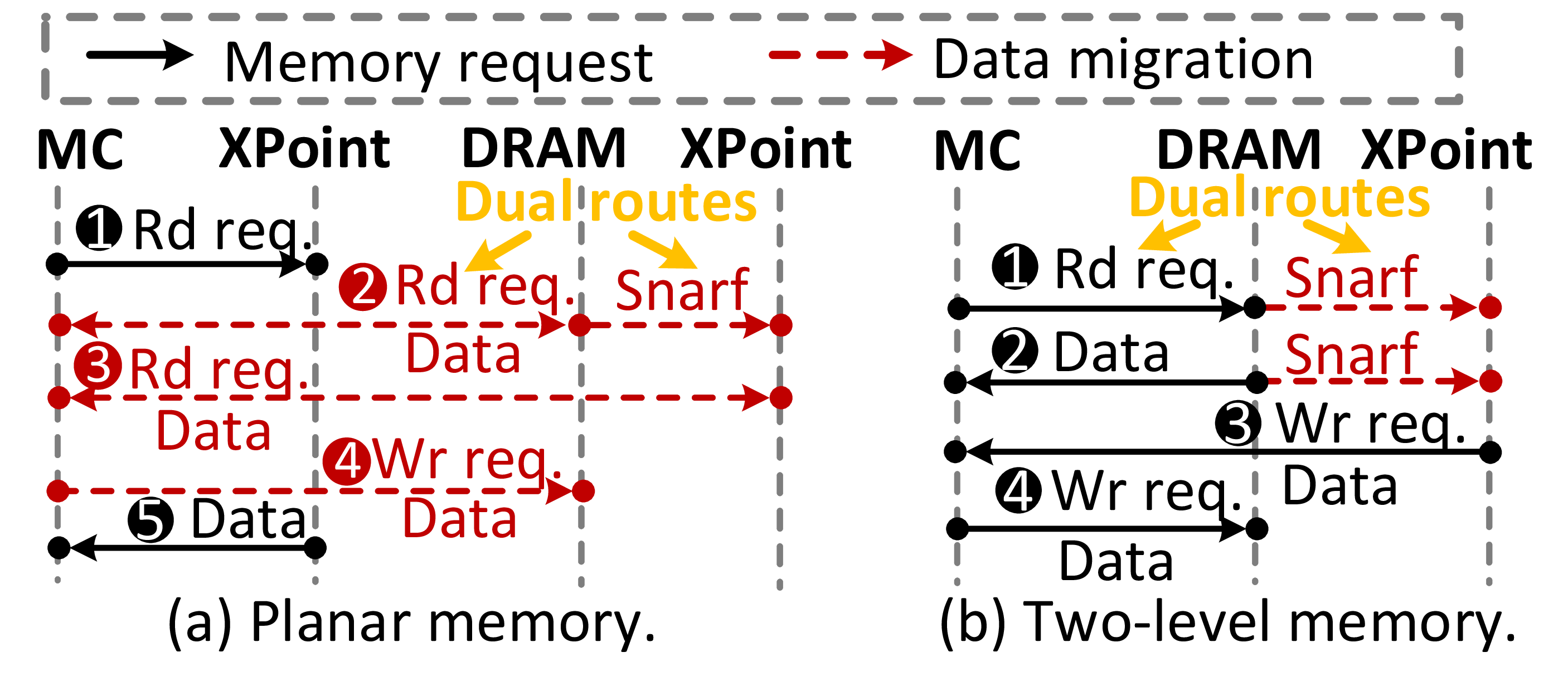}
	%\vspace{-15pt} 
	\caption{\label{fig:timing}Timing diagram of auto-read/write.\vspace{-5pt}}
	\vspace{-5pt}
\end{figure}

\noindent \textbf{Auto-read/write function.} 
Traditional memory system relies on the memory controller to copy data back and forth for data migrations between two memory modules. The data migrations make the memory channel and controller unavailable for 39\% of total execution time (cf. Figure \ref{fig:motiv1}).
To address this, Ohm-GPU introduces a new function, referred to as \textit{auto-read/write}, in the XPoint controller. 
Figure \ref{fig:timing} shows a timing diagram of data migrations with assistance of our auto-read/write function. As an example of the planar memory mode (cf. Figure \ref{fig:timing}a), the memory controller needs to perform the data migration between DRAM and XPoint (\redcircled{2}$\sim$\redcircled{4}) while serving a memory request (\whitecircled{1}\whitecircled{5}). To migrate data from DRAM to XPoint, the memory controller firstly reads the target data from DRAM via the data route (\redcircled{2}). During this time, our XPoint controller also performs a \emph{snarf} operation \cite{ramanujan2014dynamic}. The snarf enables us to monitor the communication between the memory controller and DRAM (\redcircled{2}) and hook necessary information from the memory route such as a command type, address, data, ECC, and tag bits. The XPoint controller utilizes the collected request information to perform a data migration in XPoint without assistance of the memory controller. Since DRAM lacks a controller to perform the snarf operation, the auto-read/write function cannot be used in the process of copying data blocks from XPoint to DRAM. That is, the memory controller still needs to read the target data from XPoint (\redcircled{3}) and write this data to DRAM (\redcircled{4}) one by one.
%To this end, serving the memory request (\whitecircled{5}) is postponed by the data migration. 
Figure \ref{fig:timing}b shows auto-read/write assisted data migrations in the two-level memory mode. 
When the memory controller inquires the target metadata/data from DRAM (\whitecircled{1}\whitecircled{2}), the XPoint controller leverages a snarf operation to extract the request information and the target metadata/data. The XPoint controller detects a DRAM cache miss by comparing the request's address and the tag bits stored in the DRAM cacheline. As the inquired data should be evicted to XPoint, the XPoint controller takes over the task of a DRAM data eviction from the memory controller. However, the memory controller is required copying new data from XPoint to the DRAM cache-line (\whitecircled{3}\whitecircled{4}).  
%When the memory controller inquires DRAM to serve the memory requests (\whitecircled{1}\whitecircled{2}), the XPoint controller extracts the request information and data via snarf operation. The XPoint controller leverages the extracted data to perform data migration. As there is DRAM cache miss, XPoint directly stores the evicted data without a communication of the memory controller.
%However, the memory controller needs to copy the target data from XPoint to DRAM (\whitecircled{3}\whitecircled{4}), which doubles the channel usage.

\begin{figure}
	\centering
	\includegraphics[width=1\linewidth]{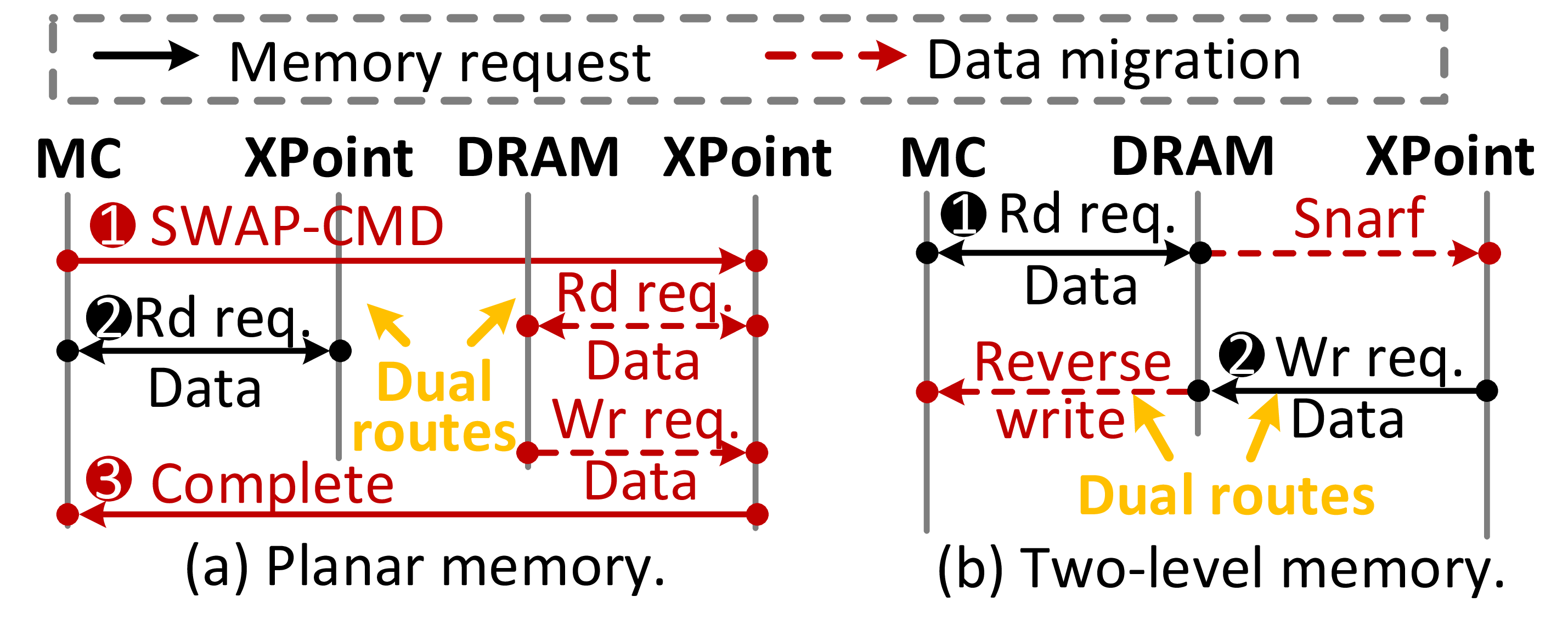}
	%\vspace{-15pt}
	\caption{\label{fig:timing_swap}Timing diagram of data migration with our auto-read/write, swap, and reverse write.\vspace{-5pt}}
	\vspace{-5pt}
\end{figure}

\noindent \textbf{Swap and reverse write functions.}
%While the auto-read/write function enables direct data transfers from DRAM to XPoint via the dual routes, moving the data in a reverse direction still requires assistance of memory controller; that is, data should be firstly moved from XPoint to the memory controller through the main data path, and memory controller also need to forwards the same data to DRAM. To make the data migration more efficient, 
To reduce the overhead of migrating data from XPoint to DRAM, we design two new memory functions, each being referred to as \emph{swap} and \emph{reverse write}. These new functions compensate the drawbacks of the auto-read/write. Specifically, the swap function allows the XPoint controller to directly manage the read and write memory transactions in DRAM via the DDR interface. However, the swap function faces three challenges: first, unlike the auto-read/write function, which snarfs I/O commands and data from the memory channel, a XPoint controller should be informed of the request information to initialize the swap function. Second, the XPoint controller cannot generate correct DDR command sequences as it is unaware of the states (i.e., precharged, activated, etc.) of the target DRAM bank. Third, the memory controller may have channel-level conflicts with the XPoint controller during the execution of our swap function. Since the memory controllers cannot issue memory requests during a DRAM refresh period, one possible solution to resolve the conflicts is to constrain the direct data copy operations in the DRAM refresh period \cite{lee2020nvdimm}. This approach, unfortunately, serializes the memory requests and data migration requests, which can severely degrade the GPU performance. To address all these challenges, we compose a new memory command, \emph{SWAP-CMD}, which is used by the memory controller to demand the swap function in the XPoint controller (cf. \redcircled{1} in Figure \ref{fig:timing_swap}a). This command reuses the data route to transfer the metadata information such as DRAM address, XPoint address, and data size. Afterwards, the memory controller can serve the incoming memory request (\whitecircled{2}) while the XPoint controller migrates data between DRAM and XPoint. The XPoint controller signals the memory controller once the data migration completes (\redcircled{3}).
Note that, since the memory controller records the states of all DRAM banks, we leverage the memory controller to preset the target DRAM bank to a stable state (i.e., activated state) before it issues the SWAP command to the XPoint controller. To avoid a potential memory channel conflict, the XPoint controller can synchronize its progress with the memory controller by leveraging the DDR-T protocol. 

The swap function yet cannot completely reduce memory traffic for the two-level memory mode due to two issues: 1) the communication between the memory controller and DRAM cannot be avoided, as the memory controller always needs to read data and tag information from DRAM to check if there is a DRAM cache miss; 2) when there is a DRAM cache miss, a memory controller should read the target data from XPoint to serve the memory request as soon as possible, rather than being stalled by the swap function. Instead, the reverse write function can address such the issue by collaborating with our auto-read/write function. Specifically, while the auto-read/write function can use the dual routes to reduce the bandwidth of transferring data from DRAM to XPoint (cf. \whitecircled{1} in Figure \ref{fig:timing_swap}b), our reverse-write function can leverage the same independent route to move data from XPoint to DRAM in addition to transferring data to the memory controller through the data route (cf. \whitecircled{2} in Figure \ref{fig:timing_swap}b). 

%\begin{figure}
%\centering
%\subfloat[The fraction of dirty blocks.]{\label{fig:dirtyblock}{\includegraphics[width=0.56\linewidth]{figs/dirtyblock}}}
%\hspace{2pt}
%\subfloat[Memory scheduling.]{\label{fig:mem_schedule}{\includegraphics[width=0.39\linewidth]{figs/mem_schedule}}}
%\caption{\label{fig:motiv_schedule}Breakdown analysis for the backup pages and memory scheduling policy for swap function.}
%\vspace{-15pt}
%\end{figure}

\subsection{Optical Network Infrastructure}
\noindent \textbf{MRR design for dual routes.}
The traditional optical channel does not allow the co-existence of dual routes owing to a design issue of the MRR-based photonic transmitter and receiver \cite{shacham2007design,beausoleil2008nanoelectronic}. In practice, the laser light of one wavelength can be modulated and absorbed by only a pair of photonic transmitter and receiver, respectively. This makes it difficult for other photonic transmitters or receivers to either reuse or snarf the same laser light. We may employ an optical power splitter \cite{vantrease2008corona} to split a single laser light to multiple memory devices such that the same light bits can be shared among the memory controller, DRAM and XPoint. However, most power splitters cannot change their status (i.e., enabled/disabled) or adjust the split ratio at a runtime. This inflexibility renders the power splitters difficult to build a flexible independent route between any two memory devices. Although \cite{zhou2013probe, morris2013extending} proposed a tunable power splitter, the tuning time, unfortunately, is as long as 6 $ns$, which is even longer than DRAM data burst latency. Instead, we adopt an idea from \cite{peter2016active}, which proposes a new tunable power splitter based on a modified MRR. Compared to the traditional power splitter \cite{vantrease2008corona,zhou2013probe, morris2013extending}, the MRR-based power splitter is flexible to change the split ratio and consumes shorter tuning time. \cite{peter2016active} reports that the MRR-based power splitter reduces the resonance tuning time to only 500 $ps$. The key insight behind this approach is that MRR can partially couple the laser light of a specific wavelength if the MRR is tuned to be in partial resonance with the laser light. The uncoupled laser light can traverse to other memory devices. The status between fully-coupled (MRR fully absorbs the laser light) and non-coupled (not absorbs the laser light) is called \emph{half-coupled}. We'll explain how a half-coupled MRR (HCMRR) enables dual routes in an optical channel, shortly.   

\noindent \textbf{Photonic transmitter and receiver design.} To enable the dual routes for our auto-read/write and reverse write functions, Ohm-GPU configures the photonic receiver in DRAM as the half-coupled MRR. Since the half-coupled MRR based photonic receiver only splits the laser light rather than fully absorbs it, the laser light, which is modulated by the memory controller, can transverse to the XPoint. The XPoint controller can, thus, snarf request/data information from the laser light. On the other hand, to enable dual routes for swap function, one laser light stream needs to carry two pieces of data, one from memory requests and another from data migration requests. To this end, we leverage a write-once memory (WOM) coding \cite{zhang2015opennvm}. By using WOM coding, Ohm-GPU can modulate two different 2-bit data in a 3-bit laser light signal. We will explain more details in Section \ref{sec:hcmrr}. Since a 3-bit laser light signal can only carry 2-bit data, WOM coding reduces the effective memory bandwidth by 33\% for the memory requests. To avoid the bandwidth wastage, we propose an aggressive approach. Specifically, photonic transmitters are also configured as the half-coupled MRR to enable the dual routes for our swap function. While the traditional photonic transmitter in the memory controller modulates the data bit 0 by fully absorbing the laser light, the half-coupled MRR ensures that the laser light maintains at least half of power strength, regardless of carrying the data bit 0 or 1. Thanks to the half-coupled MRR, the XPoint controller can reuse the remaining laser light to modulate its own data. 
%To enable dual routes for swap function, Ohm-GPU needs to employ half-tuned MRR as both photonic modulator and receiver. However, the laser light power can attenuate by upto 4 times in the receiver side, compared to the condition where swap function is disabled. Thus, the VCSEL sources need to increase their laser light power by 4 times to compensate for the power attenuation. To reduce the power consumption, we remove the usage of half-tuned MRR for photonic modulators. Instead, we leverage the write-once memory coding \cite{} to write two pieces of data on the same laser light stream. Specifically,

%\begin{figure*}
%\centering
%\subfloat[Swap function.]{\label{fig:swap}{\includegraphics[width=0.4\linewidth]{figs/swap}}}
%\hspace{2pt}
%\subfloat[Reverse-write function.]{\label{fig:reversewrite}{\includegraphics[width=0.4\linewidth]{figs/reversewrite}}}
%\hspace{2pt}
%\subfloat[Memory scheduling.]{\label{fig:mem_schedule}{\includegraphics[width=0.18\linewidth]{figs/mem_schedule}}}
%\caption{\label{fig:motiv_schedule}The design details of swap and reverse-write functions, and the memory scheduling policy for swap function.}
%\vspace{-15pt}
%\end{figure*}

\begin{figure}
	\centering
	\includegraphics[width=0.9\linewidth]{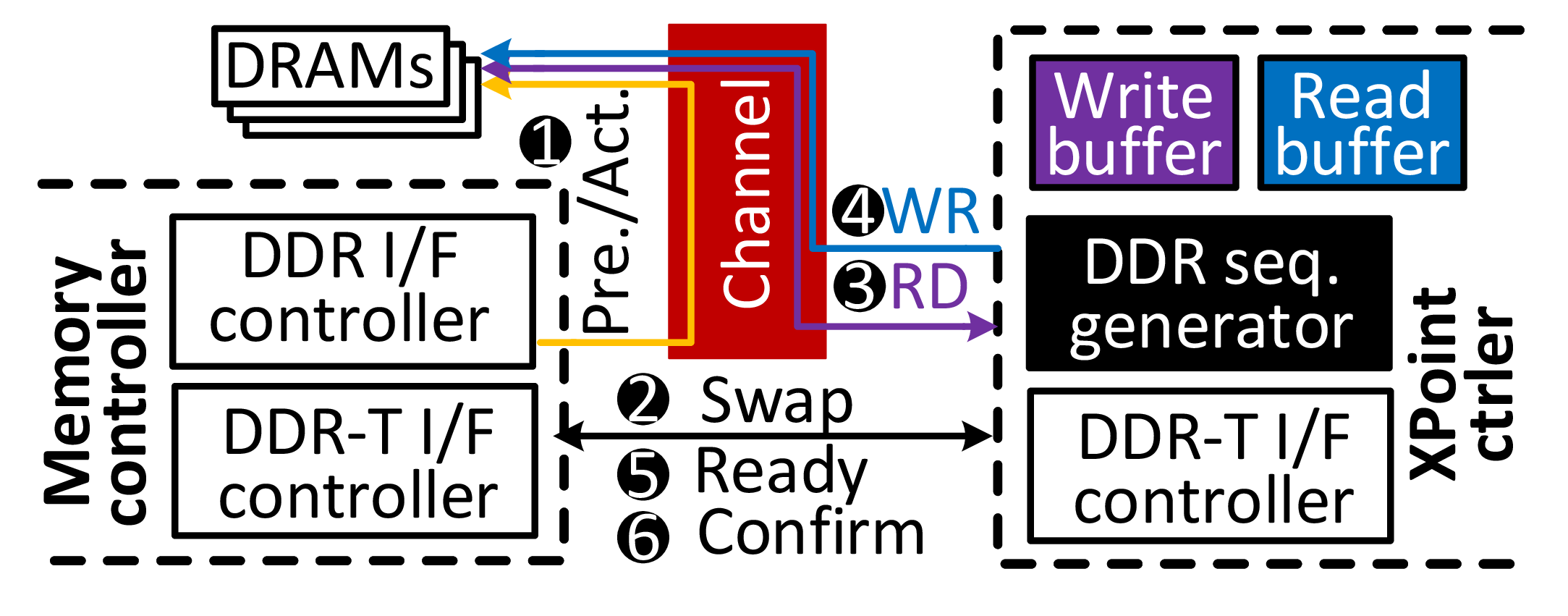}
	%\vspace{-5pt}
	\caption{\label{fig:swap}Design details of swap function.\vspace{-5pt}}
	\vspace{-5pt}
\end{figure}

\begin{figure}
	\centering
	\includegraphics[width=0.9\linewidth]{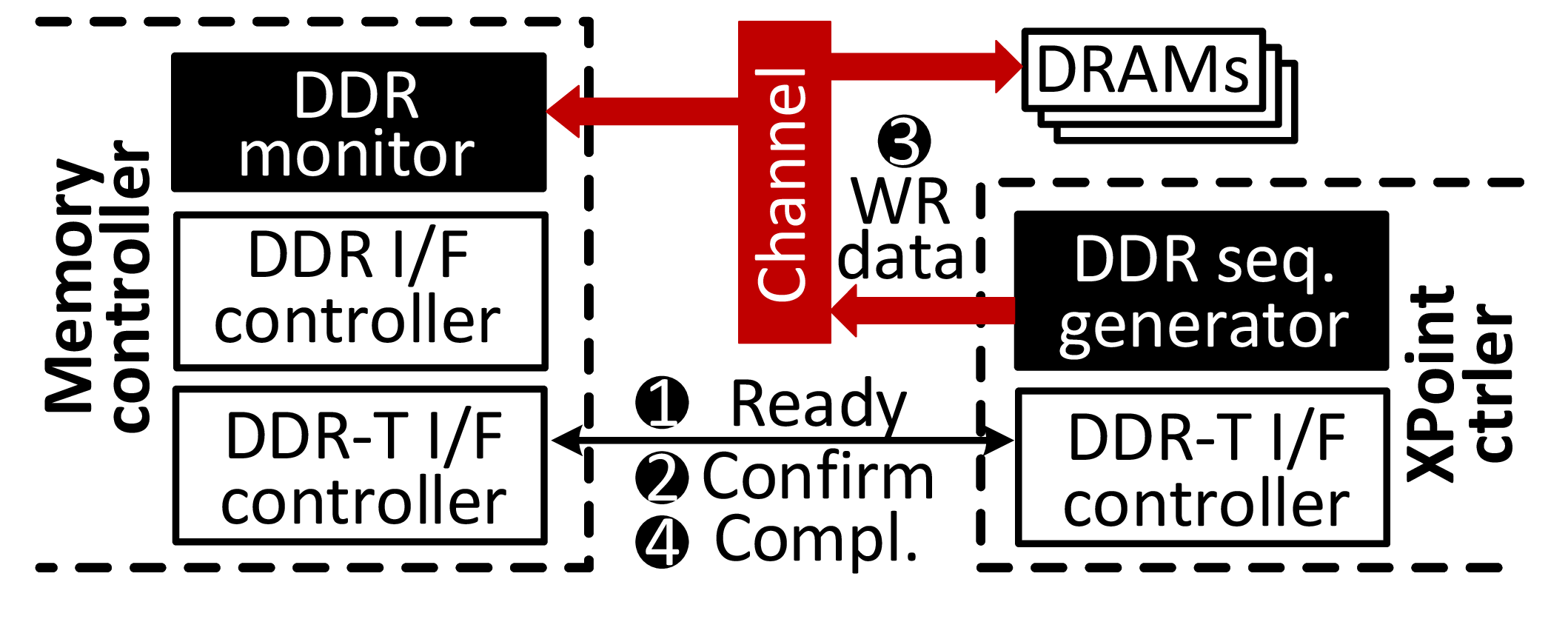}
	%\vspace{-5pt}
	\caption{\label{fig:reversewrite}Design details of our reverse-write function.\vspace{-5pt}}
	\vspace{-5pt}
\end{figure}   

\begin{figure*}
	\centering
	\includegraphics[width=1\linewidth]{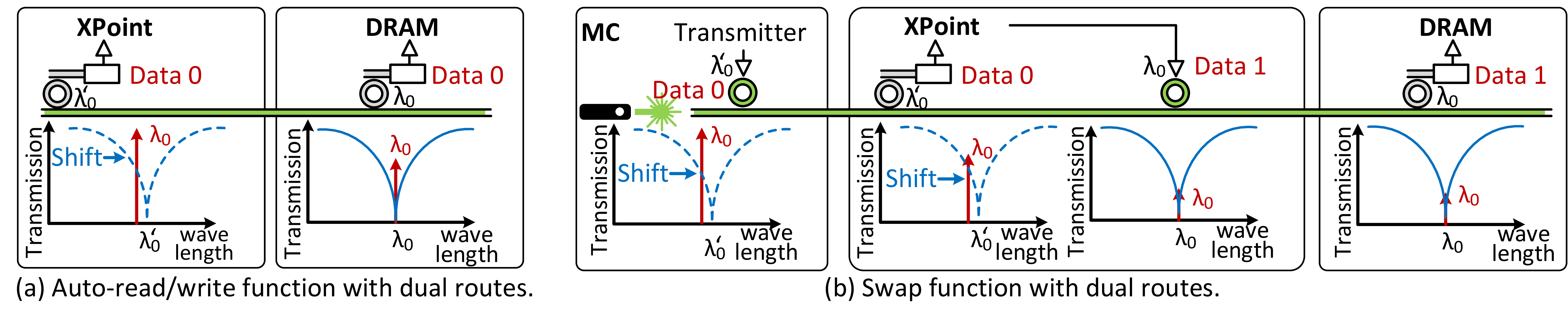}
	%\vspace{-15pt}
	\caption{\label{fig:optical}Auto-read/write and swap functions under the assistance of dual routes in the optical channel.\vspace{-5pt}}
	\vspace{-5pt}
\end{figure*}

\begin{figure}
	\centering
	\includegraphics[width=1\linewidth]{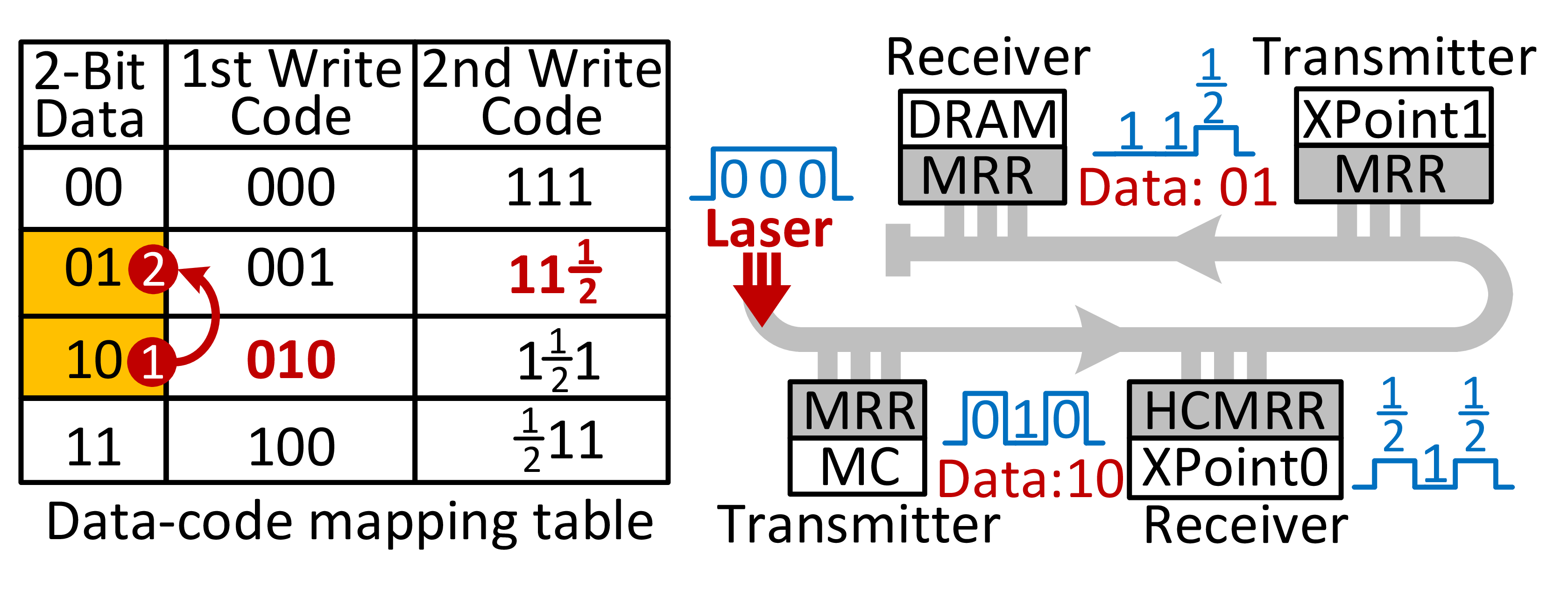}
	%\vspace{-15pt}
	\caption{\label{fig:wom}Creating dual routes with wom coding.\vspace{-5pt}}
	\vspace{-5pt}
\end{figure}

%challenge 1: the position of DRAM and PMEM is not determined. challenge 2: the number of PMEM may be greater than one.
\noindent \textbf{Optical channel optimization for low cost.} While the proposed photonic transmitter and receiver designs make the dual routes feasible in an optical channel, it is challenging to deploy the corresponding photonic hardware in our GPU memory system. The main reason behind such a difficulty is that it needs a fine-granule tuning method for one MRR to precisely switch the status among coupled, half-coupled, and non-coupled. 
%The monolithic optical pulse unfortunately cannot realize switching among three wavelength in optical-pulse tuned MRRs \cite{ibrahim2004photonic}. An alternative solution may be using electrical-tuned MRRs. However, 
For example, \cite{peter2016active} reports the tuning procedure increases the latency by 5$\times$ from 100 $ps$ to 500 $ps$, which is too long compared to the data transmission rate of the laser light.  
To address this challenge, we propose to employ an array of photonic transmitters and detectors in each memory device. To configure an independent route between any two devices, each memory device needs four pairs of photonic transmitters and detectors in total; two pairs, which are made from the fully-coupled and half-coupled MRRs, respectively, are attached to the forward path from memory controller to memory device, and the same two pairs are attached to the backward path. Unfortunately, this new design introduces a huge area cost. We optimize such array by reducing the number of MRRs to be just sufficient to serve our auto-read/write, reverse-write and swap functions. We simplify the array by leveraging the insight that different operational modes of heterogeneous memory require different memory functions. 

%\begin{figure*}
%\centering
%\subfloat[The fraction of dirty blocks.]{\label{fig:dirtyblock}\rotatebox{0}{\includegraphics[width=0.25\linewidth]{figs/dirtyblock}}}
%\hspace{2pt}
%\subfloat[Swap function.]{\label{fig:swap}\rotatebox{0}{\includegraphics[width=0.36\linewidth]{figs/swap.eps}}}
%\hspace{2pt}
%\subfloat[Reverse-write function.]{\label{fig:reversewrite}\rotatebox{0}{\includegraphics[width=0.36\linewidth]{figs/reversewrite.eps}}}
%\caption{\label{fig:motivandfunction} The fraction of clean and dirty blocks in a page for backup, and the details of swap and reverse-write functions.}
%\vspace{-15pt}
%\end{figure*}

\begin{figure*}
	\centering
	%\vspace{-10pt}
	\includegraphics[width=1\linewidth]{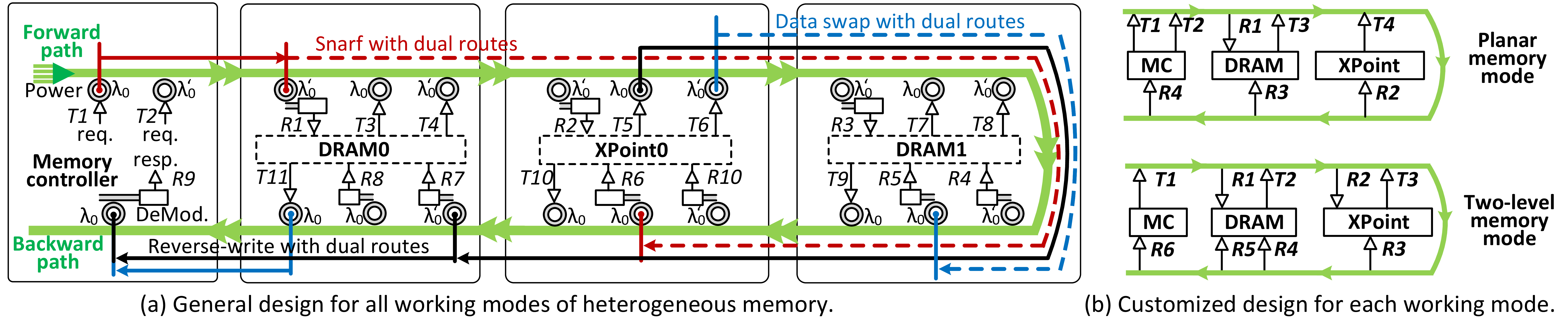}
	%\vspace{-15pt}
	\caption{\label{fig:optical_overall}The transmitter and receiver layout of our optical channel.}
	%\vspace{-5pt}
\end{figure*} 

\section{Hardware Implementation}
\label{sec:implementation}
\subsection{Swap and Reverse-Write Functions}
To realize swap function, we introduce a DDR sequence generator in the XPoint controller to manage the memory read and write transactions of DRAM.  
Figure \ref{fig:swap} shows the implementation details of our swap function. As shown in the figure, before sending a swap command to XPoint, the memory controller firstly checks the state of the target DRAM. If the target data is not loaded in DRAM row buffer, the memory controller issues the precharge and activate commands to clean the row buffer and load the target data from DRAM cells to row buffer, respectively (\whitecircled{1}). Afterwards, the memory controller informs the XPoint controller of the swap command and postpones scheduling the memory requests that conflict with the swap command (\whitecircled{2}). The XPoint controller then leverages the DDR sequence generator to control the read and write transactions of the target DRAM (\whitecircled{3}\whitecircled{4}). Once the swap function completes, the XPoint controller sends a ready signal to the memory controller via DDR-T protocol (\whitecircled{5}). The memory controller then responds with confirmation signal and resumes the stalled requests (\whitecircled{6}). 

To implement the reverse-write function, we also introduce a DDR monitor in the memory controller. Similar to XPoint, the DDR monitor module enables the memory controller to monitor the memory channel and extract request information from the communication between XPoint and DRAM. Figure \ref{fig:reversewrite} shows the implementation details of the reverse-write function. Before XPoint starts the reverse-write function, the XPoint controller sends a ready signal to notify the memory controller to snarf data from the memory channel (\whitecircled{1}). The memory controller then stops issuing new memory requests, enables the DDR monitor, and then sends a confirmation signal to the XPoint controller (\whitecircled{2}). Afterwards, the memory controller snarfs the data from the memory channel, while XPoint writes data to DRAM (\whitecircled{3}). Once XPoint sends a completion signal to the memory controller (\whitecircled{4}), the memory controller serves the requests with the collected data. 

%\subsection{Planar Memory Mode Specific Optimizations}  
%The swap function should be carefully scheduled to avoid blocking the latency-critical memory requests (i.e., read requests). We observe that memory controllers usually classify the memory requests based on the urgency and postpones the schedules of the non-urgent memory requests in the background. In practice, read requests are considered as latency-critical, while write requests are considered as non-latency-critical. Thus, we propose to schedule swap commands together with the write requests. The design details are shown in Figure \ref{fig:mem_schedule}. Specifically, we allocate a new queue, referred to as \emph{swap queue}, to store the swap commands, in addition to the read and write queues. When the read queue is empty or one of the write and swap queues is full, the memory controller starts to simultaneously drain both the write requests and swap commands. Note that the memory controller also needs to check the potential conflicts in memory channel. For example, if a swap command needs to access DRAM rank 2 and XPoint, the write requests to access either DRAM rank 2 or 3 should be forbidden from scheduling to avoid the channel conflict.

%\begin{figure*}
%	\centering
%	\includegraphics[width=1\linewidth]{figs/timing1.eps}
%	\vspace{-10pt}
%	\caption{\label{fig:timing1}Timing diagram of using phantom path.\vspace{-5pt}}
%	\vspace{-5pt}
%\end{figure*}

\subsection{Half-coupled Micro Ring Resonator}
\label{sec:hcmrr}
Figure \ref{fig:optical}a shows our approach of generating dual routes to support auto-read/write and reverse-write functions. To snarf messages from the optical channel, the photonic receiver of XPoint tunes its resonance wavelength to $\lambda_{0}^{'}$, whose frequency shifts slightly from the message-carrying wavelength $\lambda_{0}$. Due to this wavelength shift method, the photonic receiver partially couples the laser light of wavelength $\lambda_{0}$. Thus, the remaining laser light traverses from XPoint to DRAM and its data are detected by DRAM's photonic detector.

Figure \ref{fig:wom} shows how our Ohm-GPU can leverage the half-coupled MRR and WOM coding to generate our dual routes for the swap function. Specifically, by referring the data-code mapping table, the transmitters of the target memory and the XPoint controllers can encode their data into 1st and 2nd write codes, respectively. The write codes are then modulated in the same light signals, which will be detected and decoded by the corresponding receivers. In the figure, the memory and XPoint1 controllers send data to XPoint0 and DRAM, respectively. When the memory controller transfers data ``10'' to XPoint0, the target data is encoded into a write code ``010'' and modulated into a laser light (\redcircled{1}). The half-coupled MRR-based receiver (HCMRR) of XPoint0 controller gets the laser light and decodes the write code to data ``10''. This process consumes the laser light by half, which is ``$\dfrac{1}{2}$1$\dfrac{1}{2}$''. XPoint1 controller can modulate its own data ``01'' in the same laser light, which turns the light bits to ``11$\dfrac{1}{2}$'' (\redcircled{2}). DRAM can decode the laser light to retrieve data ``01'' by referring to the data-code mapping table.

To fully utilize bandwidth of the optical channel, we propose an aggressive approach to multiplex the data of memory requests and data migration requests on the same optical channel, which is shown in Figure \ref{fig:optical}b. Specifically, when modulating electrical bit `0' as light bit, the transmitter in memory controller does not tune MRR to fully couple the laser light. Instead, the photonic transmitter tunes its resonance wavelength to $\lambda_{0}^{'}$, reducing the strength of the laser light by half. For example, a data stream of ``00110'' is modulated as ``$\dfrac{1}{2}\dfrac{1}{2}$11$\dfrac{1}{2}$'' in terms of laser light strength. The photonic detector can identify bits 0 or 1 based on the light strength. Note that the photonic detector in XPoint also modulates the resonance wavelength to $\lambda_{0}^{'}$, which partially couples the laser light. 
At this juncture, the remaining laser light maintains at least one fourth of its original transmission power (``$\dfrac{1}{4}\dfrac{1}{4}\dfrac{1}{2}\dfrac{1}{2}\dfrac{1}{4}$'' in the example). The laser light continues to traverse between XPoint and DRAM.  
The transmitter of XPoint can modulate its own data stream ``10101'' in the remaining laser light by tuning MRR to fully-coupled or non-coupled. After modulation, the strength of the laser light becomes ``$\dfrac{1}{4}$0$\dfrac{1}{2}$0$\dfrac{1}{4}$''. The photonic detector in DRAM fully couples the laser light and detects the data stream of ``10101'' by checking the light strength.

\subsection{Optical Channel Optimization for Low Cost}
Creating dual routes to support auto-read/write, write-reverse and swap functions in any memory device requires at least three transmitters and three receivers in DRAM, and two transmitters and three receivers in XPoint. Figure \ref{fig:optical_overall}a shows a general optical channel design. Specifically, DRAM and XPoint need to employ a pair of MRR-based photonic transmitter and receiver (i.e., \textit{T3}, \textit{R8}, \textit{T5}, \textit{R6}, \textit{T7} and \textit{R5} in Figure \ref{fig:optical_overall}a) to support conventional photonic communication. While both DRAM and XPoint need to employ half-coupled MRR-based photonic receivers in both the forward path and backward path to enable auto-read/write function (i.e., \textit{R1}, \textit{R2}, \textit{R3}, \textit{R4}, \textit{R7} and \textit{R11}), DRAM needs half-coupled MRR-based photonic receivers in the backward path to enable write-reverse function (i.e., \textit{R4} and \textit{R7}). Lastly, to support data swap function in the dual routes, the half-coupled MRR-based photonic transmitters should be employed in DRAM and XPoint (i.e., \textit{T4}, \textit{T6} and \textit{T8}). While the photonic transmitters of \textit{T9}, \textit{T10} and \textit{T11} are not necessary, they can improve the parallelism of scheduling memory requests and data swap function in parallel. We further reduce the number of MRRs based on the characteristics of different operational modes in heterogeneous memory. Specifically, planar memory mode only needs data swap function and two-level memory mode only needs auto-read/write and reverse-write functions. Figure \ref{fig:optical_overall}b shows the MRR deployment in different operational modes. Our customized design can reduce the number of required MRRs by 58\% and 42\% in planar and two-level memory modes, respectively, compared to the general design. Note that, for simplicity, only a pair of DRAM and XPoint is depicted in the figure. But our analysis can be applied to any number of DRAM and XPoint. 

\section{Evaluation}
\label{sec:evaluation}

\noindent \textbf{Simulation methodology.}
We implement Ohm-GPU atop a GPU simulator (MacSim) \cite{kim2012macsim}. To explore a full design space of optical network based heterogeneous memory subsystems, we replace six 32-bit electrical memory channels with a single optical channel as default. This optical channel configuration can provide the same bandwidth as the traditional electrical memory channels. Note that the optical channel's bandwidth is scalable as we can increase the number of optical waveguides and wavelengths. The important parameters of our optical channel are configured based on \cite{li2013exploring}. Specifically, we adopt the static channel division policy to create six virtual channels to connect different memory devices, which matches with the number of memory controllers inside a GPU. Each virtual channel has a 16-bit data bus width and runs on 30 GHz. We then model XPoint and integrate it into MacSim. 
Its latency values are derived from the real XPoint-based NVDIMM performance measurement \cite{izraelevitz2019basic}. 
The default memory capacity of the baseline GPU is 24GB, which is the same as the memory capacity of NVIDIA K80 GPU \cite{gpuk80}.
We configure the capacity ratios of DRAM and XPoint as 1:8 and 1:64 to achieve the optimal performance \cite{optane-perf} and maximum capacity of the GPU heterogeneous memory system, respectively.
The detailed configurations are given in Table \ref{tab:sys-config}.
Considering that the simulation speed of an architectural simulator is up to 1000$\times$ slower than native-execution, we reduce the memory footprints of executed workloads to 8GB and scale down the GPU memory capacity by 12$\times$. This is a common practice in architectural studies \cite{alian2017ncap}.
%To satisfy multiple purposes, we also configure the capacities of DRAM and XPoint with different ratios for two operational modes. 

\begin{table}[]
\resizebox{\linewidth}{!}{
\begin{tabular}{|l|c|l|c|}
\hline
\multicolumn{2}{|c|}{\textbf{GPU configuration}} & \multicolumn{2}{c|}{\textbf{Memory configuration}} \\ \hline
SM/freq. & 16/1.2 GHz & tRCD (DRAM) & 25 ns \\ \hline
L1 cache & 48KB, 6-way, private & tRP (DRAM) & 10 ns \\ \hline
L2 cache & 6MB, 8-way, shared & tCL (DRAM) & 11 ns \\ \hline
Electrical channels & 6 channels/32-bit/15 GHz & PRAM read & 190 ns \\ \hline
\multicolumn{2}{|c|}{\textbf{Optical channel configuration}} & PRAM write & 763 ns \\ \hline
Channel width & 96 bits & tRRD & 5 ns \\ \hline
Frequency & 30 GHz & \multicolumn{2}{c|}{\textbf{Optical power model}} \\ \hline
Strategy & Static channel division & MRR tuning power & 200 fJ/bit \\ \hline
Virtual channel & 6 & Filter drop & 1.5 dB \\ \hline
\multicolumn{2}{|c|}{\textbf{DRAM : OP-DIMM capacity}} & Waveguide loss & 0.3 dB/cm \\ \hline
Planar memory & 1:8, 108GB & Optical splitter & 0.2 dB \\ \hline
Two-level memory & 1:64, 390GB & Detector/Modulator & 0.1/0$\sim$1 dB \\ \hline
\end{tabular}}
%\vspace{-5pt}
\caption{System configurations. \label{tab:sys-config} \vspace{-10pt}}
%\vspace{-5pt}
\end{table}

\begin{table}[]
\resizebox{\linewidth}{!}{
\begin{tabular}{|l|c|c|c|l|c|c|c|}
\hline
\textbf{Apps} & \multicolumn{1}{l|}{\textbf{APKI}} & \multicolumn{1}{l|}{\textbf{Read ratio}} & \multicolumn{1}{l|}{\textbf{Benchmark}} & \textbf{Apps} & \multicolumn{1}{l|}{\textbf{APKI}} & \multicolumn{1}{l|}{\textbf{Read ratio}} & \multicolumn{1}{l|}{\textbf{Benchmark}} \\ \hline
backp & 30 & 0.53 & \cite{che2009rodinia} & bfsdata & 84 & 0.95 & \cite{nai2015graphbig} \\ \hline
lud & 20 & 0.52 & \cite{che2009rodinia} & bfstopo & 25 & 0.97 & \cite{nai2015graphbig} \\ \hline
GRAMS & 266 & 0.7 & \cite{pouchet2012polybench} & gctopo & 93 & 0.99 & \cite{nai2015graphbig} \\ \hline
FDTD & 86 & 0.7 & \cite{pouchet2012polybench} & pagerank & 599 & 0.99 & \cite{nai2015graphbig} \\ \hline
betw & 193 & 0.99 & \cite{nai2015graphbig} & sssp & 103 & 0.98 & \cite{nai2015graphbig} \\ \hline
\end{tabular}}
%\vspace{-10pt}
\caption{Workload characteristics. \label{tab:workload-charac}}
\end{table}

\noindent \textbf{Heterogeneous memory platforms.}
We implement seven different GPU platforms: (1) \texttt{Origin}: a baseline GPU architecture with a DRAM based memory system; (2) \texttt{Hetero}: a baseline GPU architecture employing an \emph{electrical channel} integrated heterogeneous memory system. \texttt{Hetero} leverages the memory controller to migrate data between DRAM and XPoint; (3) \texttt{Ohm-base}: a baseline GPU architecture employing an \emph{optical network} integrated heterogeneous memory system; (4) \texttt{Auto-rw}: a GPU integrating the auto-read/write function into \texttt{Ohm-base}; (5) \texttt{Ohm-WOM}: a GPU platform enabling auto-read/write, reverse-write and swap functions for \texttt{Ohm-base}. This platform integrates WOM coding to enable dual routes for swap function; (6) \texttt{Ohm-BW}: compared to \texttt{Ohm-WOM}, it replaces WOM coding with half-coupled MRR-based transmitters. Lastly, we configure an \texttt{Oracle} GPU platform that employs 108GB and 390GB DRAM in planar and two-level memory modes, respectively.

%\begin{table}[]
%\resizebox{\linewidth}{!}{
%\begin{tabular}{|l|c|c|c|l|c|c|c|}
%\hline
%\multicolumn{1}{|c|}{\textbf{Apps}} & \textbf{Details} & \textbf{APKI} & \textbf{Rd ratio} & \multicolumn{1}{c|}{\textbf{Apps}} & \textbf{Details} & \textbf{APKI} & \textbf{Rd ratio} \\ \hline
%SPEC-1 & (\whitecircled{1}\whitecircled{2}\whitecircled{3}\whitecircled{4})$\times$2 & 13 & 0.64 & ycsb-a & update heavy & 2.3 & 0.47 \\ \hline
%SPEC-2 & (\whitecircled{2}\whitecircled{3}\whitecircled{5}\whitecircled{6})$\times$2 & 17 & 0.72 & ycsb-b & read intensive & 2.4 & 0.67 \\ \hline
%SPEC-3 & (\whitecircled{3}\whitecircled{6}\whitecircled{7}\whitecircled{8})$\times$2 & 10 & 0.56 & ycsb-c & read most & 2.5 & 0.68 \\ \hline
%SPEC-4 & (\whitecircled{3}\whitecircled{5}\whitecircled{7}\whitecircled{9})$\times$2 & 9 & 0.59 & ycsb-d & write  & 2.2 & 0.45 \\ \hline
%SPEC-5 & (\whitecircled{1}\whitecircled{2}\whitecircled{5}\whitecircled{9})$\times$2 & 10 & 0.62 & ycsb-f & read-update & 2.4 & 0.48 \\ \hline
%\multicolumn{8}{|c|}{Perlbench \whitecircled{1}, bzip2 \whitecircled{2}, bwaves \whitecircled{3}, mcf \whitecircled{4}, milc \whitecircled{5}, soplex \whitecircled{6}, wrf \whitecircled{7}, sjeng \whitecircled{8}, leslie\whitecircled{9}} \\ \hline
%\end{tabular}}
%\caption{Workload characteristics. \label{tab:workload-charac}\vspace{-10pt}}
%\end{table}

\noindent \textbf{Workloads and energy model.}
We select ten representative workloads from \cite{nai2015graphbig,che2009rodinia,pouchet2012polybench}, which are classified as read/write-intensive and compute/memory-intensive. The details of our evaluated workloads are shown in Table \ref{tab:workload-charac}.
%We select six representative read-intensive workloads from a graph analysis benchmark \cite{nai2015graphbig} and four representative write-intensive workloads from scientific benchmarks \cite{che2009rodinia,pouchet2012polybench}. The details of our evaluated workloads are shown in Table \ref{tab:workload-charac}.
%We generate memory-intensive workloads by mixing multiple single-thread applications from SPEC2006 benchmark \cite{henningstandard}. The number of single-thread applications is the same as the number of CPU cores in our simulation system to fully utilize the computing power. To support the system persistency mode, we adopt five different workloads from a persistent memory benchmark \cite{nalli2017analysis}. These workloads can generate enough threads to fully utilize all CPU cores. The details of our evaluated workloads are shown in Table \ref{tab:workload-charac}.
We also set up power models of DRAM, XPoint and optical channel to estimate the energy consumption of our Ohm memory system. Specifically, we adopt an empirical DRAM power model from \cite{leng2013gpuwattch} to estimate the DRAM power consumption in our simulation, while getting the average power and burst power of XPoint from \cite{optane-perf}. Our power model of the optical channel (inspired by \cite{li2013exploring}) includes MRR tuning power and power loss in each optical component (e.g., filter drop loss, waveguide loss and detector loss). The details are given in Table \ref{tab:sys-config}. 
Note that the bit error rate (BER) of an optical network is proportional to the sensing power of the photonic receiver \cite{melloni2004experimental}. To achieve a reliable end-to-end communication between the memory controllers and the DRAM/XPoint modules, we adopt the default laser light power of a single wavelength (0.73 mW) from \cite{li2013exploring}, which can guarantee bit error rate (BER) under 10$^{-15}$. To meet the reliability requirements, we also increase the laser light power of \texttt{Auto-rw}, \texttt{Ohm-WOM} and \texttt{Ohm-BW} by 2$\times$, 2$\times$ and 4$\times$, respectively.

\begin{figure}
	\centering
	%\vspace{-15pt}
	\includegraphics[width=1\linewidth]{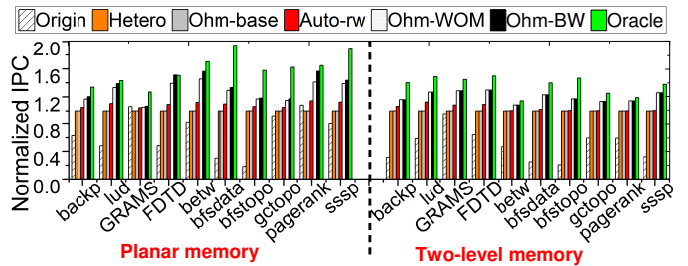}
	%\vspace{-10pt}
	\caption{\label{fig:norm_ipc}Performance of the evaluated GPU platforms.}
	%\vspace{-5pt}
\end{figure}

\begin{figure}
	\centering
	%\vspace{-10pt}
	\includegraphics[width=1\linewidth]{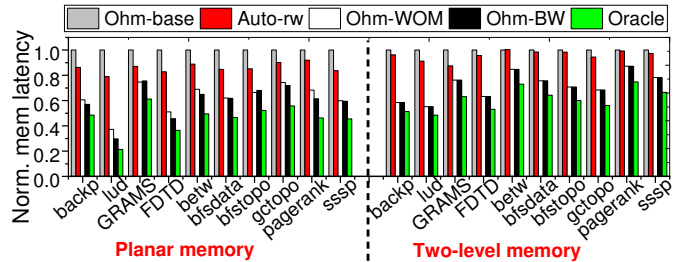}
	%\vspace{-10pt}
	\caption{\label{fig:norm_memlat}Memory access latency norm. to \texttt{Ohm-base}.}
	%\vspace{-5pt}
\end{figure}

\subsection{Overall Performance Analysis}
\label{sec:overall-eva}
\noindent \textbf{IPC.}
Figure \ref{fig:norm_ipc} shows the IPC values of different GPU platforms under various workloads, normalized to \texttt{Ohm-base}. 
%The storage capacity of \texttt{Origin}'s DRAM-based memory system is around 16$\times$ smaller than that of \texttt{Hetero}'s heterogeneous memory. 
Due to the relatively small memory size, \texttt{Origin} invokes data copies between the host and the GPU more frequently. As a result, \texttt{Origin} degrades the overall performance, compared to \texttt{Hetero}, by 42\%. Since the optical network in our default configuration has the same bandwidth as electrical memory channels (cf. Table \ref{tab:sys-config}), \texttt{Hetero} and \texttt{Ohm-base} exhibit similar performance each other. Nevertheless, \texttt{Ohm-base} consumes a lower power and less area space than \texttt{Hetero} (cf. Section \ref{sec:sensitive}). Note that \texttt{Ohm-base} can significantly outperform \texttt{Hetero} by employing multiple waveguides which will be discussed in Section \ref{sec:sensitive}. \texttt{Auto-rw} improves the performance by 9\% and 4\%, respectively, in planar and two-level memory modes, compared to \texttt{Ohm-base}. This is because \texttt{Auto-rw} leverages the auto-read/write function to automate the data transfers from DRAM to XPoint without the interference from the memory controller. 
%\texttt{Auto-rw} reduces the memory traffic generated by data migrations, which in turn mitigates the impact of data migration on the system performance. 
On the other hand, \texttt{Ohm-WOM} can improve the performance by 18\% and 16\%, compared to \texttt{Auto-rw}, in the two operational modes. Specifically, \texttt{Ohm-WOM} in planar memory mode adopts the swap function to fully decouple the memory controller from the management of data migration. \texttt{Ohm-WOM} in planar memory mode also hides the impact of data migration on latency-critical read requests by issuing the data migration with write requests at background. By adopting this approach, the performance of \texttt{Ohm-WOM} is further increased by 4\%. Lastly, \texttt{Ohm-WOM} in two-level memory mode automates the data transfer from XPoint to DRAM with the reverse-write function, which reduce the optical channel traffic. Compared to \texttt{Ohm-WOM}, \texttt{Ohm-BW} further improves the performance by 4\% in the planar mode. This is because \texttt{Ohm-BW} increases the effective bandwidth of optical channels for the memory requests, thereby improving the overall performance. 
As DRAM delivers up to 6$\times$ higher throughput than XPoint \cite{optane-perf}, \texttt{Oracle} outperforms all the evaluated GPU platforms that employ heterogeneous memory systems (i.e., \texttt{Hetero}, \texttt{Ohm-base}, \texttt{Auto-rw}, \texttt{Ohm-WOM} and \texttt{Ohm-BW}). Note that \texttt{Ohm-BW} achieves 88\% of the ideal GPU performance (\texttt{Oracle}) with a much lower cost (cf. Section \ref{sec:sensitive} for details). 
%Although DRAM delivers 6$\times$ higher throughput than XPoint, \texttt{Ohm-BW} degrades the performance by only 12\%, compared to \texttt{Oracle}. This is because \texttt{Ohm-BW} effectively leverage its DRAM cache to serve most of the incoming memory requests.

%\begin{figure*}
%\centering
%\subfloat[System persistency mode.]{\label{fig:ipc_scene1}\rotatebox{0}{\includegraphics[width=0.23\linewidth]{figs/ipc_scene1}}}
%\hspace{1pt}
%\subfloat[Planar memory mode.]{\label{fig:ipc_scene2}\rotatebox{0}{\includegraphics[width=0.37\linewidth]{figs/ipc_scene2}}}
%\hspace{1pt}
%\subfloat[Two-level memory mode.]{\label{fig:ipc_scene3}\rotatebox{0}{\includegraphics[width=0.37\linewidth]{figs/ipc_scene3}}}
%\caption{\label{fig:norm_ipc}The performance of different optic-heterogeneous memory subsystems.}
%\vspace{-15pt}
%\end{figure*}

\begin{figure}
	\centering
	%\vspace{5pt}
	\includegraphics[width=1\linewidth]{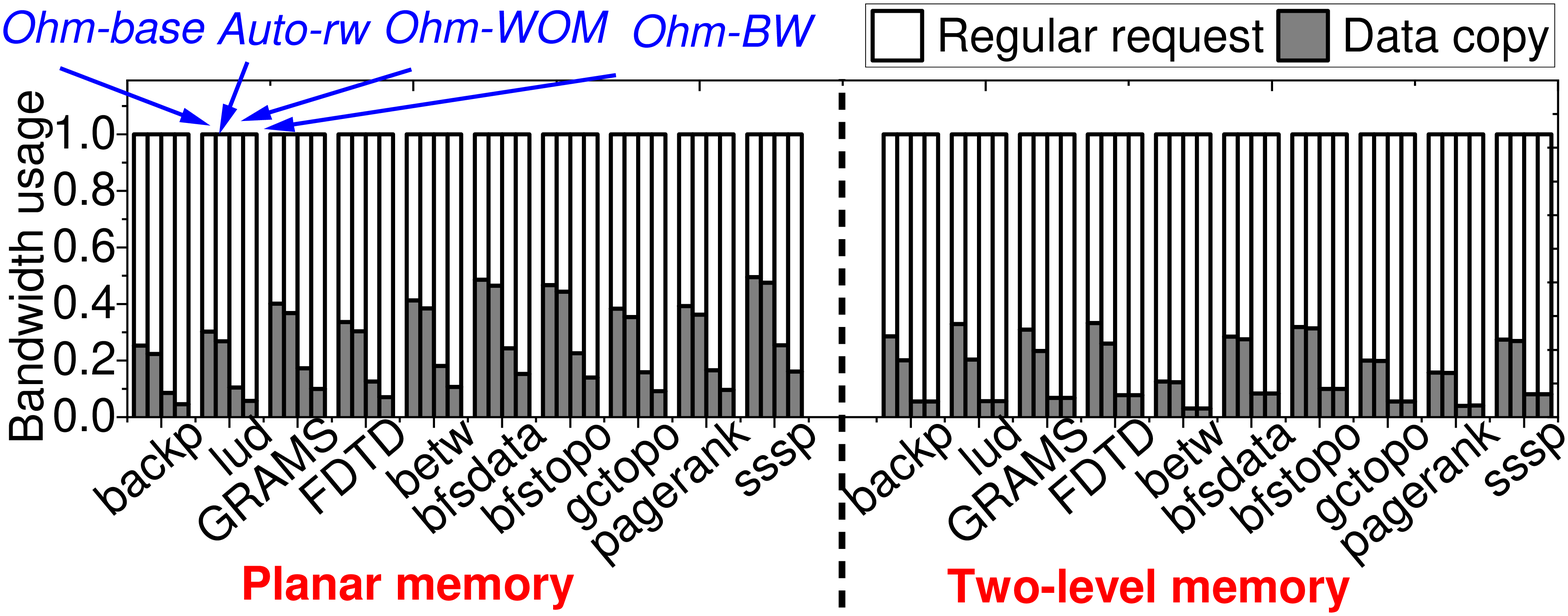}
	%\vspace{-10pt}
	\caption{\label{fig:datamove}Memory usage of evaluated GPU systems.}
	%\vspace{-5pt}
\end{figure}

\begin{figure}
	\centering
	%\vspace{-10pt}
	\includegraphics[width=1\linewidth]{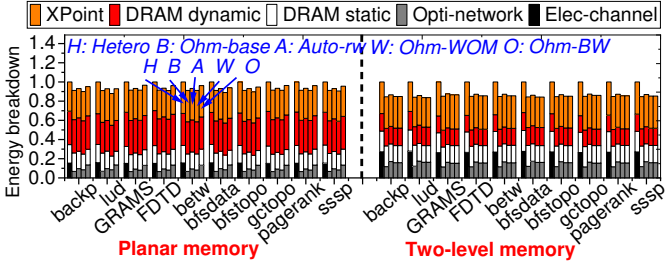}
	%\vspace{-10pt}
	\caption{\label{fig:norm_energy}Energy breakdown of GPU memory systems.}
	%\vspace{-10pt}
\end{figure}

%\begin{figure*}
%\centering
%\subfloat[System persistency mode.]{\label{fig:memlat_scene1}\rotatebox{0}{\includegraphics[width=0.23\linewidth]{figs/memlat_scene1}}}
%\hspace{1pt}
%\subfloat[Planar memory mode.]{\label{fig:memlat_scene2}\rotatebox{0}{\includegraphics[width=0.37\linewidth]{figs/memlat_scene2}}}
%\hspace{1pt}
%\subfloat[Two-level memory mode.]{\label{fig:memlat_scene3}\rotatebox{0}{\includegraphics[width=0.37\linewidth]{figs/memlat_scene3}}}
%\caption{\label{fig:norm_energy}The average memory access latency of different optic-heterogeneous memory subsystems.}
%\vspace{-15pt}
%\end{figure*}

\noindent \textbf{Memory latency analysis.}
Figure \ref{fig:norm_memlat} shows the average memory latencies of different GPU platforms under various workloads, normalized to \texttt{Ohm-base}. \texttt{Auto-rw} can reduce the average memory latencies by 14\% and 4\%, respectively, in planar and two-level memory modes, compared to \texttt{Ohm-base}. This is because \texttt{Auto-rw} can reduce the time of the optical channel to do data migration, so that memory devices can use the optical channel earlier to serve the regular memory requests. \texttt{Ohm-WOM} in planar memory mode leverages swap function to schedule the data migration through the dual routes in parallel with the memory requests, which does not occupy the optical channel. Thus, \texttt{Ohm-WOM} achieves the minimum memory access latency, which is 28\% shorter than \texttt{Auto-rw}, in planar memory mode. Similarly, the benefit of memory latency reduction brought by \texttt{Ohm-WOM} in two-level memory mode is 24\%, compared to \texttt{Auto-rw}. Although the reverse-write function in \texttt{Ohm-WOM} cannot shorten the latency of serving the memory requests from XPoint, the reverse-write function leverages the dual routes to reduce the latency of copying data from XPoint to DRAM, making the data migration procedure complete earlier. Compared to \texttt{Ohm-WOM}, \texttt{Ohm-BW} can serve the memory requests faster when swap function is invoked. Therefore, \texttt{Ohm-BW} further reduces the memory latency by 6\% in planar memory mode. 
\texttt{Ohm-BW} reduces the performance gap between a heterogeneous memory system and an ideal memory system from 67\% to 18\%. This is because \texttt{Ohm-BW} effectively leverage its DRAM cache to serve the most incoming memory requests.
%Lastly, \texttt{Oracle} achieves 15\% shorter memory latency than \texttt{Ohm-BW}. However, the benefits come from an ideal memory system without data migration and XPoint delay, which unfortunately is not practical.

%\begin{figure*}
%\centering
%\subfloat[System persistency mode.]{\label{fig:datamove_scene1}\rotatebox{0}{\includegraphics[width=0.23\linewidth]{figs/datamove_scene1}}}
%\hspace{1pt}
%\subfloat[Planar memory mode.]{\label{fig:datamove_scene2}\rotatebox{0}{\includegraphics[width=0.37\linewidth]{figs/datamove_scene2}}}
%\hspace{1pt}
%\subfloat[Two-level memory mode.]{\label{fig:datamove_scene3}\rotatebox{0}{\includegraphics[width=0.37\linewidth]{figs/datamove_scene3}}}
%\caption{\label{fig:datamove}The memory usage of different optic-heterogeneous memory subsystems.}
%\vspace{-15pt}
%\end{figure*}

%\begin{figure*}
%\centering
%\subfloat[System persistency mode.]{\label{fig:energy_scene1}\rotatebox{0}{\includegraphics[width=0.23\linewidth]{figs/energy_scene1}}}
%\hspace{1pt}
%\subfloat[Planar memory mode.]{\label{fig:energy_scene2}\rotatebox{0}{\includegraphics[width=0.37\linewidth]{figs/energy_scene2}}}
%\hspace{1pt}
%\subfloat[Two-level memory mode.]{\label{fig:energy_scene3}\rotatebox{0}{\includegraphics[width=0.37\linewidth]{figs/energy_scene3}}}
%\caption{\label{fig:norm_energy}The Energy breakdown of different optic-heterogeneous memory subsystems.}
%\vspace{-15pt}
%\end{figure*}

\noindent \textbf{Optical channel usage.}
Figure \ref{fig:datamove} shows the fraction of the optical channel bandwidth consumed by data migration in different memory platforms. As \texttt{Auto-rw} can partially do the data migration in the dual routes, it can reduce the bandwidth usage for data migration by 8\% and 17\%, respectively, in planar and two-level memory modes. \texttt{Ohm-WOM} in planar memory mode further reduces the bandwidth usage by 54\% because it performs most of data migrations in the dual routes. \texttt{Ohm-WOM} in the two-level memory mode can fully eliminate the optical channel occupancy caused by data migration. This is because the auto-read/write and reverse-write functions migrate the data in an independent route, when DRAM and XPoint serve the memory requests.

\noindent \textbf{Energy consumption.}
Figure \ref{fig:norm_energy} shows the energy consumption of different components within the memory systems of the evaluated GPU platforms. 
Compared to \texttt{Hetero}, \texttt{Ohm-base} replaces the electrical memory channels with an optical channel, which can reduce the DMA's power by 57\%, on average. 
%Thus, \texttt{Ohm-base} can cut-down the overall energy consumption by 12\%, compared to \texttt{Hetero}. 
As the user applications generate a fixed amount of memory requests to DRAM, the dynamic DRAM energy to serve the memory requests is the same as those of \texttt{Ohm-base}, \texttt{Auto-rw}, \texttt{Ohm-WOM} and \texttt{Ohm-BW}.
%On the other hand, the XPoint in different heterogeneous memory platforms serve the same number of memory requests, which consumes the same amount of energy. Therefore, the average power of XPoint is impacted by the total execution time.
On the other hand, the static DRAM energy is proportional to the total execution time when DRAM is powered on. As \texttt{Ohm-WOM} can reduce the execution time of user applications, it, on average, can reduce the static DRAM energy by 19\% and 11\%, respectively, compared to \texttt{Ohm-base} and \texttt{Auto-rw}. 
Similar to dynamic DRAM energy, the XPoint in different heterogeneous memory platforms serve the same number of memory requests, which consumes the same amount of energy. \texttt{Auto-rw}, \texttt{Ohm-WOM} and \texttt{Ohm-BW} consume higher optical channel energy than \texttt{Ohm-base}. This is because the laser source needs to increase its power to meet the sensing accuracy of the photonic detectors in the dual routes. As laser power has a minor impact on the energy cost than DRAM and XPoint, \texttt{Ohm-WOM} still decreases the overall energy consumption by 2\% and 1\% in planar and two-level memory modes, against \texttt{Ohm-base}.

\begin{figure}
\centering
%\vspace{-10pt}
\subfloat[Sensitive testing.]{\label{fig:sensitive3}\rotatebox{0}{\includegraphics[width=0.48\linewidth]{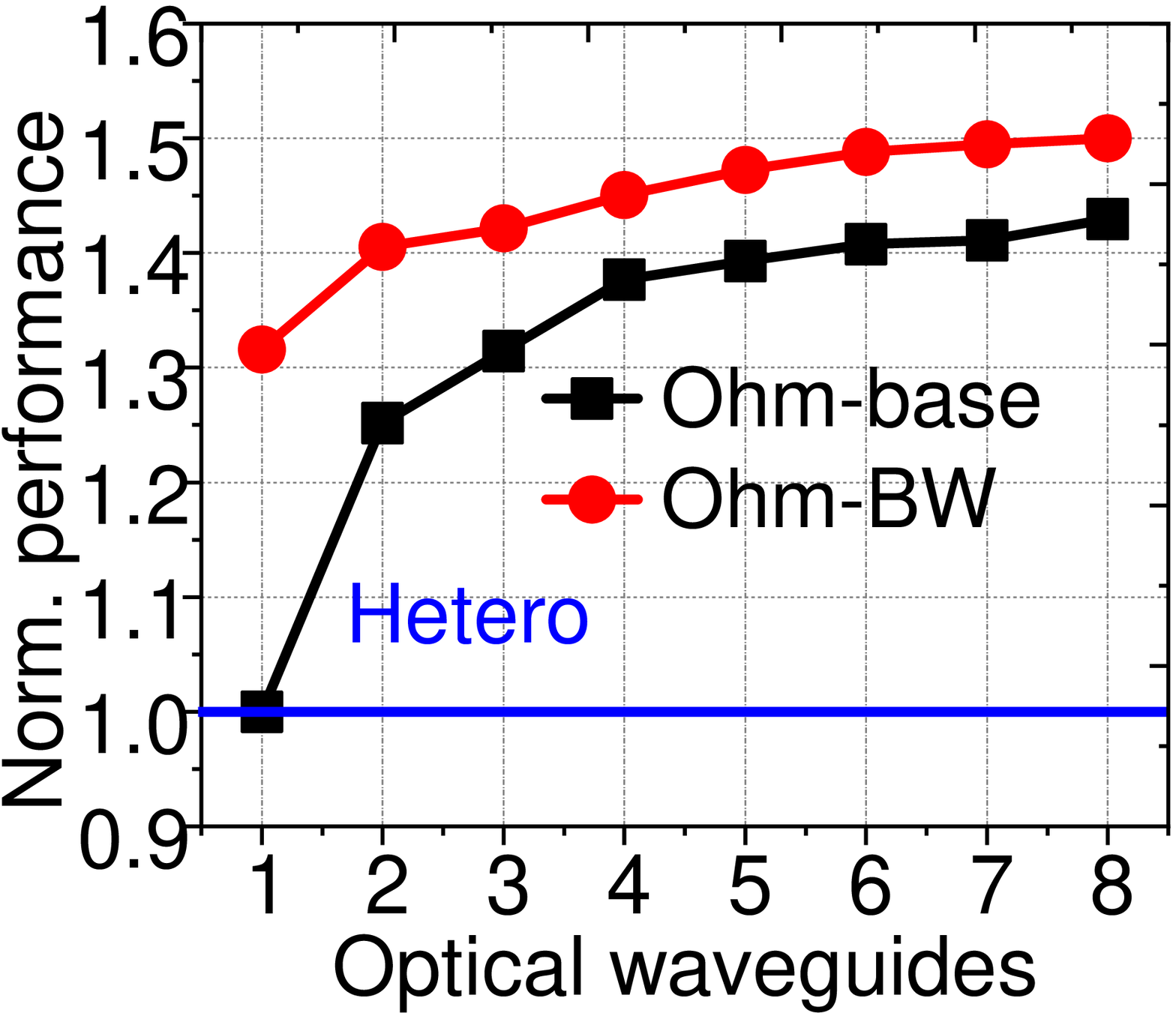}}}
\hspace{3pt}
\subfloat[Bit error rates.]{\label{fig:ber}\rotatebox{0}{\includegraphics[width=0.48\linewidth]{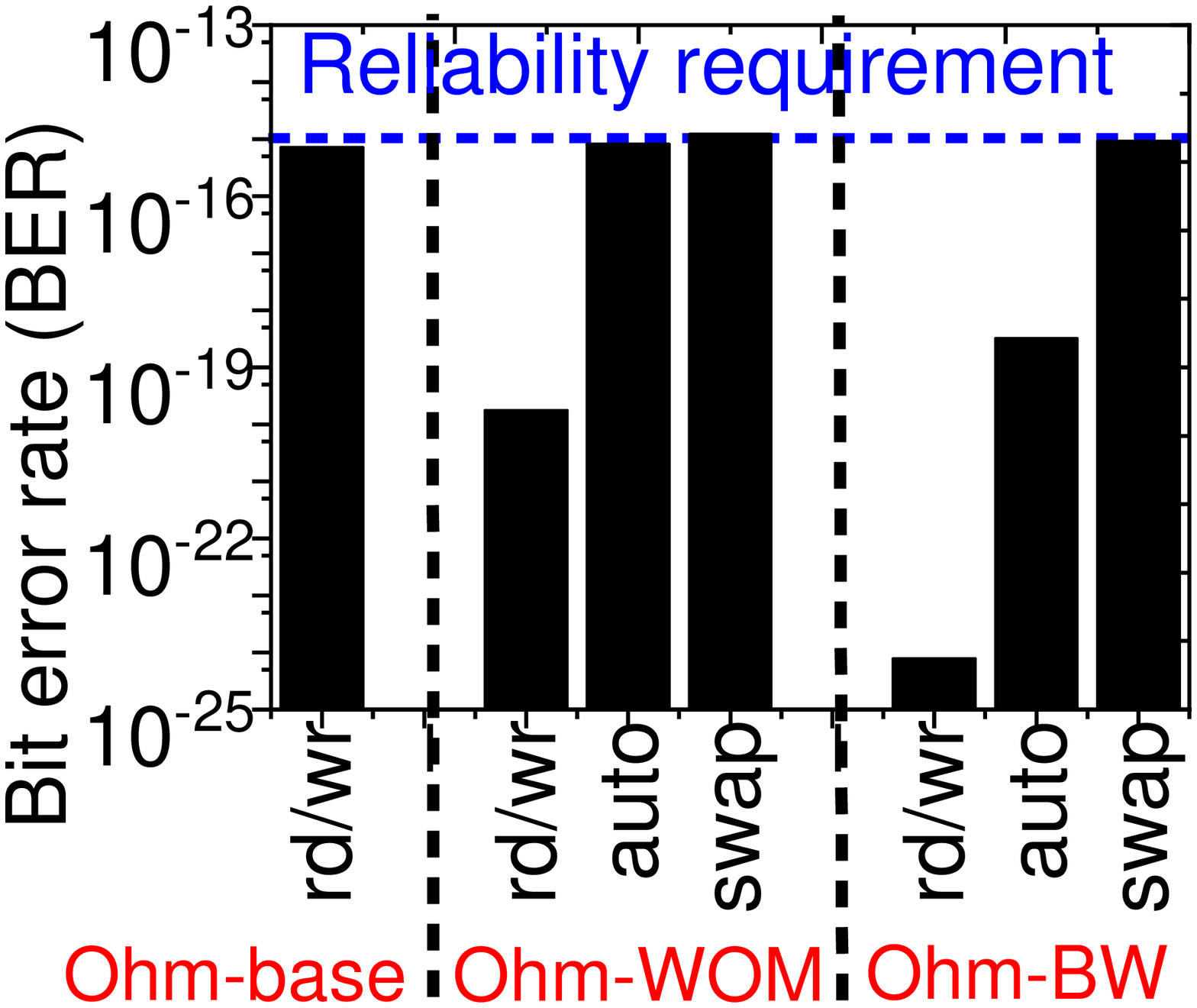}}}
%\vspace{-5pt}
\caption{\label{fig:}Performance analysis of varying optical waveguides and reliability analysis of Ohm-GPU.}
\end{figure}

%\begin{figure}
%\centering
%\subfloat[System persistency mode.]{\label{fig:writeamp1}\rotatebox{0}{\includegraphics[width=0.44\linewidth]{figs/writeamp1}}}
%\hspace{1pt}
%\subfloat[Planar memory mode.]{\label{fig:datacopy1}\rotatebox{0}{\includegraphics[width=0.44\linewidth]{figs/datacopy2}}}
%\caption{\label{fig:analysis}Write amplification analysis in system persistency mode and data migration analysis of \texttt{Phantom} in planar memory mode.}
%\vspace{-5pt}
%\end{figure}

%\begin{figure}
%	\centering
%	\includegraphics[width=1\linewidth]{figs/sensitive2.eps}
%	\vspace{-10pt}
%	\caption{\label{fig:sensitive1}Performance of \texttt{Ohm-base} with different optical channel bandwidths.\vspace{-10pt}}
%	\vspace{-10pt}
%\end{figure}

\subsection{Sensitive testing, reliability and Overhead}
\label{sec:sensitive}

\noindent \textbf{Sensitive tests.}
While a single optical waveguide can achieve the bandwidth same as electrical memory channels of 192 lanes, Ohm-GPU can employ multiple optical waveguides under the same area constrains as the electrical memory channels. Figure \ref{fig:sensitive3} shows the performance improvement brought by multiple optical waveguides in Ohm-GPU. \texttt{Ohm-base} with 8 optical waveguides improves the system performance than \texttt{Hetero}, by 41\%, on average. This is because employing multiple optical waveguides can significantly reduce the DMA latency of the heterogeneous memory system. \texttt{Ohm-BW} also benefits from the increased number of optical waveguides. The performance improvement can be 17\%.

\noindent \textbf{Reliability.} 
We evaluate the data integrity of the end-to-end communication in various GPU memory systems, and the results are shown in Figure \ref{fig:ber}. 
%A reliable optical channel is expected to have BER under 10$^{-15}$ \cite{}.
When serving memory requests, \texttt{Ohm-base} can achieve BER as low as 7.2$\times$10$^{-16}$ with our default configuration of laser light power. Since \texttt{Ohm-WOM} and \texttt{Ohm-BW} deploy more MRR-based modulators and detectors in the optical infrastructure for the auto-read/write and swap functions, modulating and detecting data in these MRRs can attenuate the strength of laser signals (cf. optical power model in Table \ref{tab:sys-config}). To compensate for the laser power loss, we increase the powers of the laser source by 2$\times$ and 4$\times$ in \texttt{Ohm-WOM} and \texttt{Ohm-BW}, respectively. The measured BER of auto-read/write and swap functions are 6.1$\times$10$^{-16}$ and 9.9$\times$10$^{-16}$, respectively, in \texttt{Ohm-WOM}, while the worst BER in \texttt{Ohm-BW} is 9.3$\times$10$^{-16}$. To sum up, the optical channel of Ohm-GPU satisfies the reliability requirement of 10$^{-15}$ BER.

\begin{table}[]
%\vspace{-10pt}
\resizebox{\linewidth}{!}{
\begin{tabular}{|l|c|c|c|c|}
\hline
\textbf{Modes} & \multicolumn{2}{c|}{\textbf{Planar memory}} & \multicolumn{2}{c|}{\textbf{Two-level memory}} \\ \hline
\textbf{DRAM} & \multicolumn{2}{c|}{1GB x 12, \$140} & \multicolumn{2}{c|}{1GB x 6, \$70} \\ \hline
\textbf{XPoint} & \multicolumn{2}{c|}{8GB x 12, \$125} & \multicolumn{2}{c|}{32GB x 12, \$499} \\ \hline
 & Modulators & Detectors & Modulators & Detectors \\ \hline
\textbf{Ohm-base} & 2,112/\$3 & 2,112/\$3 & 2,368/\$3 & 2,368/\$3 \\ \hline
%\textbf{Auto-rw} & 2,368 & 2,880 & 2,112 & 4,160 & 2,368 & 2,880 \\ \hline
%\textbf{All-support} & 6,272/\newedit{\$9} & 6,208/\newedit{\$9} & 7,040/\newedit{\$10} & 6,976/\newedit{\$10} \\ \hline
\textbf{Ohm-BW} & 2,176/\$3 & 3,136/\$5 & 2,368/\$3 & 4,928/\$7 \\ \hline
\textbf{VCSEL} & \multicolumn{4}{c|}{\$100} \\ \hline
\end{tabular}}
%\vspace{-5pt}
\caption{\label{fig:overhead}Cost estimation of different Ohm memories. \vspace{-10pt}}
\end{table}

\begin{figure}
	\centering
	%\vspace{-10pt}
	\includegraphics[width=1\linewidth]{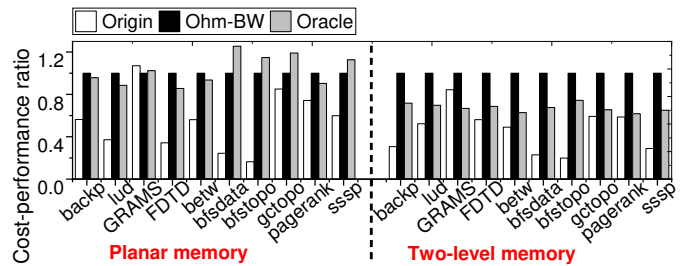}
	%\vspace{-10pt}
	\caption{\label{fig:cpvalue_fig}Cost-performance analysis of \texttt{Origin}, \texttt{Ohm-BW} and \texttt{Oracle} GPU platforms (higher is better).}
	%\vspace{-5pt}
\end{figure}

\noindent \textbf{Overhead analysis.}
Table \ref{fig:overhead} lists the estimated cost of the key components in different platforms. We configure up to 24 memory devices in a GPU, while the number of DRAM and XPoint chips are set to follow the capacity ratio listed in Table \ref{tab:sys-config}. The price of memory devices are calculated based on \cite{gddr6-price,optane-price}. We refer to Figure \ref{fig:optical_overall} to calculate the total number of MRRs required in each memory platform. 
%Compared to \texttt{Ohm-base}, \texttt{All-support} increases the total number of MRRs by 168\% to support all proposed hardware functions.
%\texttt{Ohm-BW} decreases the total number of MRRs by 63\%, 57\% and 48\%, respectively, compared to \texttt{All-support}. 
The fabrication cost of MRRs is estimated based on \cite{optic-fab}.
Although \texttt{Ohm-BW} employs 41\% more MRRs (at the cost of \$4) than \texttt{Ohm-base} to support the dual routes in the optical channel, \texttt{Ohm-BW} improves the performance by 20\%, demonstrating higher performance/cost ratio. 
%Considering that \texttt{Ohm-BW} reduce the requirement of the optical resource by leveraging dual routes to address data migration issue, we believe it is worthwhile to implement dual routes in the optical channel.
In addition to optical components, \texttt{Ohm-BW} also employs a DDR monitor and DDR sequence generator in the memory controller and the XPoint controller, respectively. The DDR monitor is comprised of control logic and a few registers, which enable/disable a DDR interface controller to receive data from the data bus, while the DDR sequence generator converts the memory requests into DDR packets. Our implementation of DDR sequence generators requires 2.8K LUTs and 4.7K flip-flops in FPGA, whose cost is negligible to the XPoint controller.
Considering the launch price of NVIDIA K80 GPU (\$5k), planar and two-level memory modes enabled \texttt{Ohm-BW} only increase total cost by 7.6\% and 13.5\%, respectively.
We also evaluate the cost-performance (CP) ratio of NVIDIA K80 GPU (\texttt{Origin}), our Ohm-GPU (\texttt{Ohm-BW}) and an ideal GPU (\texttt{Oracle}), and the results are shown in Figure \ref{fig:cpvalue_fig}. The CP ratio of \texttt{Ohm-BW} is 155\% and 24\% higher than that of \texttt{Origin} and \texttt{Oracle}, respectively, indicating that the performance benefits brought by Ohm-GPU overwhelms its cost overhead.
%which makes Ohm-GPU a promising product for customers.
%Although \texttt{Ohm-BW} costs more than \texttt{Origin}, \texttt{Ohm-BW} increases the CP ratio by 155\%, on average, compared to \texttt{Origin}.

%\begin{table}[]
%\vspace{5pt}
%\resizebox{\linewidth}{!}{
%\begin{tabular}{|l|c|c|c|c|c|c|}
%\hline
%\textbf{Modes} & \multicolumn{2}{c|}{\textbf{System persistency}} & \multicolumn{2}{c|}{\textbf{Planar memory}} & \multicolumn{2}{c|}{\textbf{Two-level memory}} \\ \hline
%\textbf{DRAM} & \multicolumn{2}{c|}{32GB x 32} & \multicolumn{2}{c|}{32GB x 16} & \multicolumn{2}{c|}{32GB x 4} \\ \hline
%\textbf{PMEM} & \multicolumn{2}{c|}{256GB x 4} & \multicolumn{2}{c|}{256GB x 16} & \multicolumn{2}{c|}{256GB x 32} \\ \hline
% & Modulators & Detectors & Modulators & Detectors & Modulators & Detectors \\ \hline
%\textbf{Hetero} & 2,368 & 2,368 & 2,112 & 2,112 & 2,368 & 2,368 \\ \hline
%%\textbf{Auto-rw} & 2,368 & 2,880 & 2,112 & 4,160 & 2,368 & 2,880 \\ \hline
%\textbf{All-support} & 7,040 & 6,976 & 6,272 & 6,208 & 7,040 & 6,976 \\ \hline
%\textbf{Phantom} & 2,368 & 2,880 & 2,176 & 3,136 & 2,368 & 4,928 \\ \hline
%\end{tabular}}
%\caption{\label{fig:overhead}The cost estimation of photonic modulators and detectors for the Ohm memories.}
%\vspace{-5pt}
%\end{table}

\section{Related Work}
\label{sec:relatedwork}

\noindent \textbf{Optical network.}
To address an expensive energy cost and limited scalability of electrical wires, multiple prior work \cite{goswami2012integrating, goswami2014exploring, ziabari2015leveraging, vantrease2008corona, hadke2008ocdimm, beamer2010re, li2013exploring} have proposed to integrate optical network into the memory system as a replacement of electrical wires. Specifically, \cite{goswami2012integrating, goswami2014exploring, ziabari2015leveraging} propose to substitute the traditional electrical network-on-chip with optical on-chip crossbar interconnects in a 3D-stacked GPU, which can improve throughput of interconnect network while saving its operating power. On the other hand,\cite{vantrease2008corona} configures the optical network for both inter-core and off-chip memory communication, which can easily achieve 20 TB/s bandwidth with lower power. In addition to the benefits of bandwidth and power, \cite{hadke2008ocdimm} reports a significant network latency reduction and memory capacity expansion by replacing the store-and-forward network and protocol in FB-DIMM with the low-latency optical network. \cite{beamer2010re} and \cite{li2013exploring} further optimize the optical memory subsystem by revising either the memory architecture or the optical network. \cite{beamer2010re} proposes to reduce the number of activated bits in each DRAM bank to reduce power consumption, while \cite{li2013exploring} proposes to dynamically assign multiple wavelengths as virtual channels to connect to different memory devices, which can mitigate the overhead of rank-to-rank switch delay. While the prior research achieves great progress in integrating the optical network in a homogeneous memory, none of them consider using optical network to address the challenges of a heterogeneous memory. To the best of our knowledge, Ohm-GPU is the first to address the data migration issue in the heterogeneous memory by leveraging the unique characteristics of optical network. To mitigate the impact of data migration on optical network, Ohm-GPU modified the designs of both optical network and heterogeneous memory architecture.

\noindent \textbf{Heterogeneous memory.}
There are many prior studies \cite{baek2012dual, ham2013disintegrated, wang2016duang, hwang2013hmmsched, kwon2012case} oriented towards achieving the balance of high performance and large capacity in the memory system by employing both fast-but-small memory and large-but-slow memory. \cite{baek2012dual} proposes to group DRAM and PRAM as a heterogeneous memory, while \cite{wang2016duang} places SLC PRAM and MLC PRAM together. However, all these studies adopt a legacy PRAM model, which uses a low-power protocol (LPDDR2-NVM) and exhibits a read/write latency close to DRAM latency. Recently, \cite{optane-perf} and \cite{patil2019performance} reveal the performance of combining DRAM and XPoint-based Optane DC PMM in a real computing environment. However, their research is limited to use Optane DC PMM as a black box without digging deeper into the internal architecture. In contrast, we revealed the key component designs of the heterogeneous memory. We also explored the design space of XPoint to collaborate with optical network to address the data migration issue.

\noindent \textbf{Data migration.}
There also exist many studies that try to mitigate the impact of data migration on the memory channel. \cite{seshadri2013rowclone} proposes to leverage the data bus within each DRAM rank to perform intra-bank or inter-bank data migration. While data migration in this approach does not occupy the memory channel, the data migration is constrained to operate within the DRAM rank. Although \cite{wang2016duang} allows data migration across two PRAM ranks in a lightweight way, the proposed approach relies on a unique architecture where the two PRAM ranks share the same row buffers. While the techniques proposed in both prior work are designed for homogeneous memory, our Ohm-GPU makes it feasible to mitigate the impact of data migration across different DIMMs by leveraging the dual routes in the optical network.

\section{Conclusion}
\label{sec:conclusion}
%Graphics processing units (GPUs) have been widely adopted as an efficient accelerator hardware platform to speed up the execution of large-scale data-intensive applications. However, GPUs suffer from the low memory capacity and demand for high memory bandwidth.
In this work, we propose Ohm-GPU, a new design of optical network for heterogeneous memory integrated GPU, which can mitigate the impact of data migration on optical channel. Specifically, Ohm-GPU decouples the memory controller from the management of the data migration and leverages the dual routes in optical channel to prevent data migration from occupying the memory channel. 
%Our Ohm-GPU exhibits 181\% better performance than baseline memory system.
Our Ohm-GPU can improve the performance by 181\% and 27\%, compared to a DRAM-based GPU memory system and the baseline optical network based heterogeneous memory system, respectively.
%Our evaluation indicates that Phantom saves the optical channel bandwidth by XX\%, and thus exhibits XX\% better performance than baseline optic-heterogeneous memory.

\section{Acknowledgement}
%\begin{acks}
This research is mainly supported by NRF 2021R1AC4001773 and IITP 2021-0-00524. The work is also supported in part by KAIST start-up package (G01190015), and Samsung (G01200447). Other product names used in this publication are for identification purposes only and may be trademarks of their respective companies. Myoungsoo Jung is the corresponding author.
%\end{acks}

%%%%%%% -- PAPER CONTENT ENDS -- %%%%%%%%

%%%%%%%%% -- BIB STYLE AND FILE -- %%%%%%%%
\bibliographystyle{IEEEtranS}
\bibliography{ohmgpu}

% Generated by IEEEtranS.bst, version: 1.14 (2015/08/26)
\begin{thebibliography}{10}
\providecommand{\url}[1]{#1}
\csname url@samestyle\endcsname
\providecommand{\newblock}{\relax}
\providecommand{\bibinfo}[2]{#2}
\providecommand{\BIBentrySTDinterwordspacing}{\spaceskip=0pt\relax}
\providecommand{\BIBentryALTinterwordstretchfactor}{4}
\providecommand{\BIBentryALTinterwordspacing}{\spaceskip=\fontdimen2\font plus
\BIBentryALTinterwordstretchfactor\fontdimen3\font minus
  \fontdimen4\font\relax}
\providecommand{\BIBforeignlanguage}[2]{{%
\expandafter\ifx\csname l@#1\endcsname\relax
\typeout{** WARNING: IEEEtranS.bst: No hyphenation pattern has been}%
\typeout{** loaded for the language `#1'. Using the pattern for}%
\typeout{** the default language instead.}%
\else
\language=\csname l@#1\endcsname
\fi
#2}}
\providecommand{\BIBdecl}{\relax}
\BIBdecl

\bibitem{alian2017ncap}
M.~Alian, A.~H. Abulila, L.~Jindal, D.~Kim, and N.~S. Kim, ``Ncap:
  Network-driven, packet context-aware power management for client-server
  architecture,'' in \emph{2017 IEEE International Symposium on High
  Performance Computer Architecture (HPCA)}.\hskip 1em plus 0.5em minus
  0.4em\relax IEEE, 2017, pp. 25--36.

\bibitem{hbm2}
AMD, ``High-bandwidth memory (hbm),''
  \emph{\url{https://www.amd.com/system/files/documents/high-bandwidth-memory-hbm.pdf}},
  2015.

\bibitem{aoyama20033}
M.~Aoyama, K.~Ogasawara, M.~Sugawara, T.~Ishibashi, T.~Ishibashi, S.~Shimoyama,
  K.~Yamaguchi, T.~Yanagita, and T.~Noma, ``3 gbps, 5000 ppm spread spectrum
  serdes phy with frequency tracking phase interpolator for serial ata,'' in
  \emph{2003 Symposium on VLSI Circuits. Digest of Technical Papers (IEEE Cat.
  No. 03CH37408)}.\hskip 1em plus 0.5em minus 0.4em\relax IEEE, 2003, pp.
  107--110.

\bibitem{baek2012dual}
S.~Baek, H.~G. Lee, C.~Nicopoulos, and J.~Kim, ``A dual-phase compression
  mechanism for hybrid dram/pcm main memory architectures,'' in
  \emph{Proceedings of the Great lakes symposium on VLSI}.\hskip 1em plus 0.5em
  minus 0.4em\relax ACM, 2012, pp. 345--350.

\bibitem{bahadori2016comprehensive}
M.~Bahadori, S.~Rumley, D.~Nikolova, and K.~Bergman, ``Comprehensive design
  space exploration of silicon photonic interconnects,'' \emph{Journal of
  Lightwave Technology}, vol.~34, no.~12, pp. 2975--2987, 2016.

\bibitem{beamer2010re}
S.~Beamer, C.~Sun, Y.-J. Kwon, A.~Joshi, C.~Batten, V.~Stojanovi{\'c}, and
  K.~Asanovi{\'c}, ``Re-architecting dram memory systems with monolithically
  integrated silicon photonics,'' in \emph{ACM SIGARCH Computer Architecture
  News}, vol.~38, no.~3.\hskip 1em plus 0.5em minus 0.4em\relax ACM, 2010, pp.
  129--140.

\bibitem{beausoleil2008nanoelectronic}
R.~G. Beausoleil, P.~J. Kuekes, G.~S. Snider, S.-Y. Wang, and R.~S. Williams,
  ``Nanoelectronic and nanophotonic interconnect,'' \emph{Proceedings of the
  IEEE}, vol.~96, no.~2, pp. 230--247, 2008.

\bibitem{bergman2014photonic}
K.~Bergman, L.~P. Carloni, A.~Biberman, J.~Chan, and G.~Hendry, \emph{Photonic
  network-on-chip design}.\hskip 1em plus 0.5em minus 0.4em\relax Springer,
  2014.

\bibitem{chang2007efficient}
L.-P. Chang, ``On efficient wear leveling for large-scale flash-memory storage
  systems,'' in \emph{Proceedings of the 2007 ACM symposium on Applied
  computing}, 2007, pp. 1126--1130.

\bibitem{che2009rodinia}
S.~Che, M.~Boyer, J.~Meng, D.~Tarjan, J.~W. Sheaffer, S.-H. Lee, and
  K.~Skadron, ``Rodinia: A benchmark suite for heterogeneous computing,'' in
  \emph{2009 IEEE international symposium on workload characterization
  (IISWC)}.\hskip 1em plus 0.5em minus 0.4em\relax Ieee, 2009, pp. 44--54.

\bibitem{memorywall1}
N.~Dahad, ``Addressing memory wall is key to edge-based ai,''
  \emph{\url{https://www.eetimes.com/document.asp?doc_id=1333456}}, 2018.

\bibitem{dhiman2009pdram}
G.~Dhiman, R.~Ayoub, and T.~Rosing, ``Pdram: A hybrid pram and dram main memory
  system,'' in \emph{2009 46th ACM/IEEE Design Automation Conference}.\hskip
  1em plus 0.5em minus 0.4em\relax IEEE, 2009, pp. 664--669.

\bibitem{memorywall2}
M.~Feldman, ``Revisiting the memory wall,''
  \emph{\url{https://www.hpcwire.com/2009/02/19/revisiting_the_memory_wall/}},
  2009.

\bibitem{gai2016cost}
K.~Gai, M.~Qiu, and H.~Zhao, ``Cost-aware multimedia data allocation for
  heterogeneous memory using genetic algorithm in cloud computing,'' \emph{IEEE
  transactions on cloud computing}, 2016.

\bibitem{goswami2014exploring}
N.~Goswami, Z.~Li, R.~Shankar, and T.~Li, ``Exploring silicon nanophotonics in
  throughput architecture,'' \emph{IEEE Design \& Test}, vol.~31, no.~5, pp.
  18--27, 2014.

\bibitem{goswami2012integrating}
N.~Goswami, Z.~Li, A.~Verma, R.~Shankar, and T.~Li, ``Integrating nanophotonics
  in gpu microarchitecture,'' in \emph{Proceedings of the 21st international
  conference on Parallel architectures and compilation techniques}, 2012, pp.
  425--426.

\bibitem{gupta2009dftl}
A.~Gupta, Y.~Kim, and B.~Urgaonkar, ``Dftl: a flash translation layer employing
  demand-based selective caching of page-level address mappings,'' \emph{Acm
  Sigplan Notices}, vol.~44, no.~3, pp. 229--240, 2009.

\bibitem{hadke2008ocdimm}
A.~Hadke, T.~Benavides, S.~B. Yoo, R.~Amirtharajah, and V.~Akella, ``Ocdimm:
  Scaling the dram memory wall using wdm based optical interconnects,'' in
  \emph{2008 16th IEEE Symposium on High Performance Interconnects}.\hskip 1em
  plus 0.5em minus 0.4em\relax IEEE, 2008, pp. 57--63.

\bibitem{gddr6-price}
H.~Hagedoorn, ``Gddr6 significantly more expensive than gddr5,''
  \emph{\url{https://www.guru3d.com/news-story/gddr6-significantly-more-expensive-than-gddr5.html}},
  2019.

\bibitem{ham2013disintegrated}
T.~J. Ham, B.~K. Chelepalli, N.~Xue, and B.~C. Lee, ``Disintegrated control for
  energy-efficient and heterogeneous memory systems,'' in \emph{2013 IEEE 19th
  International Symposium on High Performance Computer Architecture
  (HPCA)}.\hskip 1em plus 0.5em minus 0.4em\relax IEEE, 2013, pp. 424--435.

\bibitem{hbm2vsgddr6}
Hamza, ``Hbm2 vs gddr6 memory comparison,''
  \emph{\url{https://flashbang.biz/hbm2-vs-gddr6-memory-specification-and-full-comparison/}},
  2019.

\bibitem{optic-fab}
T.~Hausken, ``Breaking the cost barrier,''
  \emph{\url{https://www.osa-opn.org/home/articles/volume_26/july_august_2015/departments/breaking_the_cost_barrier/}},
  2015.

\bibitem{optane-size}
J.~Hruska, ``Optane dc persistent memory offers up to 512gb per dimm,''
  \emph{\url{https://www.extremetech.com/computing/288854-optane-dc-persistent-memory-offers-up-to-512gb-per-dimm}},
  2019.

\bibitem{hwang2013hmmsched}
W.~Hwang and K.~H. Park, ``Hmmsched: Hybrid main memory-aware task scheduling
  on multicore systems,'' in \emph{Proceedings of international conference on
  future computational technologies and applications}, vol.~52, 2013.

\bibitem{optaneDC}
Intel, ``Intel optane dc persistent memory,''
  \emph{\url{https://www.intel.com/content/www/us/en/architecture-and-technology/optane-dc-persistent-memory.html}},
  2018.

\bibitem{overclock}
------, ``How to overclock ram,''
  \emph{\url{https://www.intel.com/content/www/us/en/gaming/resources/overclock-ram.html}},
  2019.

\bibitem{izraelevitz2019basic}
J.~Izraelevitz, J.~Yang, L.~Zhang, J.~Kim, X.~Liu, A.~Memaripour, Y.~J. Soh,
  Z.~Wang, Y.~Xu, S.~R. Dulloor \emph{et~al.}, ``Basic performance measurements
  of the intel optane dc persistent memory module,'' \emph{arXiv preprint
  arXiv:1903.05714}, 2019.

\bibitem{optane-perf}
J.~Izraelevitz, J.~Yang, L.~Zhang, J.~Kim, X.~Liu, A.~Memaripour, Y.~J. Soh,
  Z.~Wang, Y.~Xu, S.~R. Dulloor, J.~Zhao, and S.~Swanson, ``Basic performance
  measurements of the intel optane dc persistent memory module,''
  \emph{\url{https://arxiv.org/abs/1903.05714}}, 2019.

\bibitem{kang2014co}
U.~Kang, H.-S. Yu, C.~Park, H.~Zheng, J.~Halbert, K.~Bains, S.~Jang, and J.~S.
  Choi, ``Co-architecting controllers and dram to enhance dram process
  scaling,'' in \emph{The memory forum}, vol.~14, 2014.

\bibitem{optane-test}
P.~Kennedy, ``A close look at intel optane dc persistent memory modules,''
  \emph{\url{https://www.servethehome.com/a-close-look-at-intel-optane-dc-persistent-memory-modules/}},
  2018.

\bibitem{kim2012macsim}
H.~Kim, J.~Lee, N.~B. Lakshminarayana, J.~Sim, J.~Lim, and T.~Pho, ``Macsim: A
  cpu-gpu heterogeneous simulation framework user guide,'' \emph{Georgia
  Institute of Technology}, 2012.

\bibitem{kim2002space}
J.~Kim, J.~M. Kim, S.~H. Noh, S.~L. Min, and Y.~Cho, ``A space-efficient flash
  translation layer for compactflash systems,'' \emph{IEEE Transactions on
  Consumer Electronics}, vol.~48, no.~2, pp. 366--375, 2002.

\bibitem{kim2015architectural}
Y.~Kim, ``Architectural techniques to enhance dram scaling,'' Ph.D.
  dissertation, figshare, 2015.

\bibitem{kim2014flipping}
Y.~Kim, R.~Daly, J.~Kim, C.~Fallin, J.~H. Lee, D.~Lee, C.~Wilkerson, K.~Lai,
  and O.~Mutlu, ``Flipping bits in memory without accessing them: An
  experimental study of dram disturbance errors,'' in \emph{ACM SIGARCH
  Computer Architecture News}, vol.~42, no.~3.\hskip 1em plus 0.5em minus
  0.4em\relax IEEE Press, 2014, pp. 361--372.

\bibitem{kwon2012case}
S.~Kwon, D.~Kim, Y.~Kim, S.~Yoo, and S.~Lee, ``A case study on the application
  of real phase-change ram to main memory subsystem,'' in \emph{2012 Design,
  Automation \& Test in Europe Conference \& Exhibition (DATE)}.\hskip 1em plus
  0.5em minus 0.4em\relax IEEE, 2012, pp. 264--267.

\bibitem{lee2020nvdimm}
C.~Lee, W.~Shin, D.~J. Kim, Y.~Yu, S.-J. Kim, T.~Ko, D.~Seo, J.~Park, K.~Lee,
  S.~Choi \emph{et~al.}, ``Nvdimm-c: A byte-addressable non-volatile memory
  module for compatibility with standard ddr memory interfaces,'' in \emph{2020
  IEEE International Symposium on High Performance Computer Architecture
  (HPCA)}.\hskip 1em plus 0.5em minus 0.4em\relax IEEE, 2020, pp. 502--514.

\bibitem{leng2013gpuwattch}
J.~Leng, T.~Hetherington, A.~ElTantawy, S.~Gilani, N.~S. Kim, T.~M. Aamodt, and
  V.~J. Reddi, ``Gpuwattch: enabling energy optimizations in gpgpus,''
  \emph{ACM SIGARCH Computer Architecture News}, vol.~41, no.~3, pp. 487--498,
  2013.

\bibitem{li2013exploring}
Z.~Li, R.~Zhou, and T.~Li, ``Exploring high-performance and energy proportional
  interface for phase change memory systems,'' in \emph{2013 IEEE 19th
  International Symposium on High Performance Computer Architecture
  (HPCA)}.\hskip 1em plus 0.5em minus 0.4em\relax IEEE, 2013, pp. 210--221.

\bibitem{melloni2004experimental}
A.~Melloni, M.~Martinelli, G.~Cusmai, and R.~Siano, ``Experimental evaluation
  of ring resonator filters impact on the bit error rate in non return to zero
  transmission systems,'' \emph{Optics communications}, vol. 234, no. 1-6, pp.
  211--216, 2004.

\bibitem{micheloni2013inside}
R.~Micheloni, A.~Marelli, and K.~Eshghi, \emph{Inside solid state drives
  (SSDs)}.\hskip 1em plus 0.5em minus 0.4em\relax Springer, 2013.

\bibitem{morris2013extending}
R.~Morris, E.~Jolley, and A.~K. Kodi, ``Extending the performance and
  energy-efficiency of shared memory multicores with nanophotonic technology,''
  \emph{IEEE Transactions on Parallel and Distributed Systems}, vol.~25, no.~1,
  pp. 83--92, 2013.

\bibitem{nai2015graphbig}
L.~Nai, Y.~Xia, I.~G. Tanase, H.~Kim, and C.-Y. Lin, ``Graphbig: understanding
  graph computing in the context of industrial solutions,'' in \emph{SC'15:
  Proceedings of the International Conference for High Performance Computing,
  Networking, Storage and Analysis}.\hskip 1em plus 0.5em minus 0.4em\relax
  IEEE, 2015, pp. 1--12.

\bibitem{nair2013archshield}
P.~J. Nair, D.-H. Kim, and M.~K. Qureshi, ``Archshield: Architectural framework
  for assisting dram scaling by tolerating high error rates,'' in \emph{ACM
  SIGARCH Computer Architecture News}, vol.~41, no.~3.\hskip 1em plus 0.5em
  minus 0.4em\relax ACM, 2013, pp. 72--83.

\bibitem{nale2019memory}
B.~Nale, R.~K. Ramanujan, M.~P. SWAMINATHAN, T.~Thomas, and T.~Polepeddi,
  ``Memory channel that supports near memory and far memory access,'' May~7
  2019, uS Patent 10,282,322.

\bibitem{nvidia2012nvidia}
NIVIDIA, ``Nvidia kepler next generation cuda compute architecture,''
  \emph{Computer system}, vol.~26, pp. 63--72, 2012.

\bibitem{gv100}
Nvidia, ``Nvidia tesla v100 gpu architecture,''
  \emph{\url{http://images.nvidia.com/content/volta-architecture/pdf/volta-architecture-whitepaper.pdf}},
  2017.

\bibitem{turingGPU}
NVIDIA, ``Nvidia turing gpu architecture,''
  \emph{\url{https://www.nvidia.com/content/dam/en-zz/Solutions/design-visualization/technologies/turing-architecture/NVIDIA-Turing-Architecture-Whitepaper.pdf}},
  2019.

\bibitem{gpuk80}
Nvidia, ``Nvidia tesla k80: The world most popular gpu,''
  \emph{\url{https://www.nvidia.com/en-gb/data-center/tesla-k80/}}, 2021.

\bibitem{nvidia2009nvidia}
C.~Nvidia, ``Nvidia’s next generation cuda compute architecture: Fermi,''
  \emph{Comput. Syst}, vol.~26, pp. 63--72, 2009.

\bibitem{park2018bibim}
G.~Park, M.~Kwon, P.~Mahapatra, M.~Swift, and M.~Jung, ``$\{$BIBIM$\}$: A
  prototype multi-partition aware heterogeneous new memory,'' in \emph{10th
  $\{$USENIX$\}$ Workshop on Hot Topics in Storage and File Systems (HotStorage
  18)}, 2018.

\bibitem{patil2019performance}
O.~Patil, L.~Ionkov, J.~Lee, F.~Mueller, and M.~Lang, ``Performance
  characterization of a dram-nvm hybrid memory architecture for hpc
  applications using intel optane dc persistent memory modules,'' in
  \emph{Proceedings of the International Symposium on Memory Systems}.\hskip
  1em plus 0.5em minus 0.4em\relax ACM, 2019, pp. 288--303.

\bibitem{pawlowski2011hybrid}
J.~T. Pawlowski, ``Hybrid memory cube (hmc),'' in \emph{2011 IEEE Hot chips 23
  symposium (HCS)}.\hskip 1em plus 0.5em minus 0.4em\relax IEEE, 2011, pp.
  1--24.

\bibitem{peter2016active}
E.~Peter, A.~Thomas, A.~Dhawan, and S.~R. Sarangi, ``Active microring based
  tunable optical power splitters,'' \emph{Optics Communications}, vol. 359,
  pp. 311--315, 2016.

\bibitem{pouchet2012polybench}
L.-N. Pouchet, ``Polybench: The polyhedral benchmark suite,'' \emph{URL:
  http://www. cs. ucla. edu/pouchet/software/polybench}, 2012.

\bibitem{qureshi2009enhancing}
M.~K. Qureshi, J.~Karidis, M.~Franceschini, V.~Srinivasan, L.~Lastras, and
  B.~Abali, ``Enhancing lifetime and security of pcm-based main memory with
  start-gap wear leveling,'' in \emph{2009 42nd Annual IEEE/ACM international
  symposium on microarchitecture (MICRO)}.\hskip 1em plus 0.5em minus
  0.4em\relax IEEE, 2009, pp. 14--23.

\bibitem{ramanujan2014dynamic}
R.~K. Ramanujan, G.~J. Hinton, and D.~J. Zimmerman, ``Dynamic partial power
  down of memory-side cache in a 2-level memory hierarchy,'' Oct.~9 2014, uS
  Patent App. 13/994,726.

\bibitem{ZSSD}
Samsung, ``Ultra-low latency with samsung z-nand ssd,''
  \url{Ultra-Low_Latency_with_Samsung_Z-NAND_SSD-0.pdf}, 2017.

\bibitem{seshadri2013rowclone}
V.~Seshadri, Y.~Kim, C.~Fallin, D.~Lee, R.~Ausavarungnirun, G.~Pekhimenko,
  Y.~Luo, O.~Mutlu, P.~B. Gibbons, M.~A. Kozuch \emph{et~al.}, ``Rowclone: fast
  and energy-efficient in-dram bulk data copy and initialization,'' in
  \emph{Proceedings of the 46th Annual IEEE/ACM International Symposium on
  Microarchitecture}.\hskip 1em plus 0.5em minus 0.4em\relax ACM, 2013, pp.
  185--197.

\bibitem{shacham2007design}
A.~Shacham, K.~Bergman, and L.~P. Carloni, ``On the design of a photonic
  network-on-chip,'' in \emph{First International Symposium on Networks-on-Chip
  (NOCS'07)}.\hskip 1em plus 0.5em minus 0.4em\relax IEEE, 2007, pp. 53--64.

\bibitem{shacham2008photonic}
------, ``Photonic networks-on-chip for future generations of chip
  multiprocessors,'' \emph{IEEE Transactions on Computers}, vol.~57, no.~9, pp.
  1246--1260, 2008.

\bibitem{shulaker2014monolithic}
M.~M. Shulaker, T.~F. Wu, A.~Pal, L.~Zhao, Y.~Nishi, K.~Saraswat, H.-S.~P.
  Wong, and S.~Mitra, ``Monolithic 3d integration of logic and memory: Carbon
  nanotube fets, resistive ram, and silicon fets,'' in \emph{2014 IEEE
  International Electron Devices Meeting}.\hskip 1em plus 0.5em minus
  0.4em\relax IEEE, 2014, pp. 27--4.

\bibitem{optane-price}
B.~Tallis, ``The intel optane ssd 900p 480gb review: Diving deeper into 3d
  xpoint,''
  \emph{\url{https://www.anandtech.com/show/12136/the-intel-optane-ssd-900p-480gb-review}},
  2017.

\bibitem{siliconvia}
T.~Thomas, ``Gddr6 vs hbm2 memory,''
  \emph{\url{https://www.techsiting.com/gddr6-vs-hbm2/}}, 2019.

\bibitem{vantrease2008corona}
D.~Vantrease, R.~Schreiber, M.~Monchiero, M.~McLaren, N.~P. Jouppi,
  M.~Fiorentino, A.~Davis, N.~Binkert, R.~G. Beausoleil, and J.~H. Ahn,
  ``Corona: System implications of emerging nanophotonic technology,'' in
  \emph{ACM SIGARCH Computer Architecture News}, vol.~36, no.~3.\hskip 1em plus
  0.5em minus 0.4em\relax IEEE Computer Society, 2008, pp. 153--164.

\bibitem{wang2016duang}
H.~Wang, J.~Zhang, S.~Shridhar, G.~Park, M.~Jung, and N.~S. Kim, ``Duang: Fast
  and lightweight page migration in asymmetric memory systems,'' in \emph{2016
  IEEE International Symposium on High Performance Computer Architecture
  (HPCA)}.\hskip 1em plus 0.5em minus 0.4em\relax IEEE, 2016, pp. 481--493.

\bibitem{yang2020empirical}
J.~Yang, J.~Kim, M.~Hoseinzadeh, J.~Izraelevitz, and S.~Swanson, ``An empirical
  guide to the behavior and use of scalable persistent memory,'' in \emph{18th
  $\{$USENIX$\}$ Conference on File and Storage Technologies ($\{$FAST$\}$
  20)}, 2020, pp. 169--182.

\bibitem{yoon2013flash}
J.~H. Yoon, H.~C. Hunter, and G.~A. Tressler, ``Flash \& dram si scaling
  challenges, emerging non-volatile memory technology enablement-implications
  to enterprise storage and server compute systems,'' \emph{Flash Memory
  Summit}, 2013.

\bibitem{zhang2015nvmmu}
J.~Zhang, D.~Donofrio, J.~Shalf, M.~T. Kandemir, and M.~Jung, ``Nvmmu: A
  non-volatile memory management unit for heterogeneous gpu-ssd
  architectures,'' in \emph{2015 International Conference on Parallel
  Architecture and Compilation (PACT)}.\hskip 1em plus 0.5em minus 0.4em\relax
  IEEE, 2015, pp. 13--24.

\bibitem{zhang2020zng}
J.~Zhang and M.~Jung, ``Zng: Architecting gpu multi-processors with new flash
  for scalable data analysis,'' in \emph{2020 ACM/IEEE 47th Annual
  International Symposium on Computer Architecture (ISCA)}.\hskip 1em plus
  0.5em minus 0.4em\relax IEEE, 2020, pp. 1064--1075.

\bibitem{zhang2019flashgpu}
J.~Zhang, M.~Kwon, H.~Kim, H.~Kim, and M.~Jung, ``Flashgpu: Placing new flash
  next to gpu cores,'' in \emph{2019 56th ACM/IEEE Design Automation Conference
  (DAC)}.\hskip 1em plus 0.5em minus 0.4em\relax IEEE, 2019, pp. 1--6.

\bibitem{zhang2015opennvm}
J.~Zhang, G.~Park, M.~M. Shihab, D.~Donofrio, J.~Shalf, and M.~Jung, ``Opennvm:
  An open-sourced fpga-based nvm controller for low level memory
  characterization,'' in \emph{2015 33rd IEEE International Conference on
  Computer Design (ICCD)}.\hskip 1em plus 0.5em minus 0.4em\relax IEEE, 2015,
  pp. 666--673.

\bibitem{zhou2013probe}
L.~Zhou and A.~K. Kodi, ``Probe: Prediction-based optical bandwidth scaling for
  energy-efficient nocs,'' in \emph{2013 Seventh IEEE/ACM International
  Symposium on Networks-on-Chip (NoCS)}.\hskip 1em plus 0.5em minus 0.4em\relax
  IEEE, 2013, pp. 1--8.

\bibitem{zhou2009durable}
P.~Zhou, B.~Zhao, J.~Yang, and Y.~Zhang, ``A durable and energy efficient main
  memory using phase change memory technology,'' \emph{ACM SIGARCH computer
  architecture news}, vol.~37, no.~3, pp. 14--23, 2009.

\bibitem{zhu2013accelerating}
Q.~Zhu, T.~Graf, H.~E. Sumbul, L.~Pileggi, and F.~Franchetti, ``Accelerating
  sparse matrix-matrix multiplication with 3d-stacked logic-in-memory
  hardware,'' in \emph{2013 IEEE High Performance Extreme Computing Conference
  (HPEC)}.\hskip 1em plus 0.5em minus 0.4em\relax IEEE, 2013, pp. 1--6.

\bibitem{ziabari2015leveraging}
A.~K.~K. Ziabari, J.~L. Abell{\'a}n, R.~Ubal, C.~Chen, A.~Joshi, and D.~Kaeli,
  ``Leveraging silicon-photonic noc for designing scalable gpus,'' in
  \emph{Proceedings of the 29th ACM on International Conference on
  Supercomputing}, 2015, pp. 273--282.

\end{thebibliography}
%%%%%%%%%%%%%%%%%%%%%%%%%%%%%%%%%%%%

\end{document}